\documentclass[prl,twocolumn,showpacs,preprintnumbers,amsmath,amssymb,superscriptaddress,footinbib]{revtex4-1}

\pdfoutput=1

\usepackage[export]{adjustbox}
\usepackage{amsfonts,amsmath,times,mathrsfs,graphicx,dcolumn,bm,wasysym,xcolor,epsfig,amssymb,lipsum,color,epstopdf,bbold}
\usepackage[makeroom]{cancel}
\usepackage[english]{babel}
\usepackage[T1]{fontenc}
\usepackage[colorlinks,bookmarks=true,citecolor=red,linkcolor=blue,urlcolor=blue]{hyperref}
\usepackage{hyperref}
\usepackage[tight, FIGTOPCAP, hang, raggedright, nooneline]{subfigure}

 \definecolor{Green}{RGB}{80,182,0} 
\begin{document}

\title{Anyonic self-induced disorder in a stabilizer code:\\
quasi-many body localization in a translational invariant model}

\author{H. Yarloo}
\affiliation{Department of Physics, Sharif University of Technology, P.O.Box 
11155-9161, Tehran, Iran}
\author{A. Langari}
\email{langari@sharif.edu}
\affiliation{Department of Physics, Sharif University of Technology, P.O.Box 
11155-9161, Tehran, Iran}
\affiliation{Center of excellence in Complex Systems and Condensed Matter 
(CSCM), Sharif University of Technology, Tehran 14588-89694, Iran}
\author{A. Vaezi}
\affiliation{Department of Physics, Stanford University, Stanford, CA 94305, USA}

\begin{abstract}

We enquire into the quasi-many-body localization in topologically ordered states 
of matter, revolving around the case of Kitaev toric code on the ladder geometry, where 
different types of anyonic defects carry different masses induced by environmental errors.
Our study verifies that the presence of anyons generates
a complex energy landscape solely through braiding statistics, which suffices
to suppress the diffusion of defects in such clean, multi-component anyonic liquid. 
This non-ergodic dynamics suggests a promising scenario for 
investigation of quasi-many-body localization. 
Computing standard diagnostics evidences that a typical initial inhomogeneity of anyons gives 
birth to a glassy dynamics with an exponentially diverging time scale of the full relaxation. 
Our results unveil how self-generated disorder ameliorates the 
vulnerability of topological order away from equilibrium.
This setting provides a new platform which paves the way toward 
impeding logical errors by self-localization of anyons in a generic, high energy state, 
originated exclusively in their exotic statistics.
\end{abstract}

\pacs{75.10.Jm, 03.75.Kk, 05.70.Ln, 72.15.Rn}

\maketitle

Many-body localization (MBL)~\cite{Gornyi:2005,Basko:2006,Oganesyan:2007,Pal:2010,Bauer:2013,
Imbrie:2016} generalizes the concept of single particle localization~\cite{Anderson:1958}
to isolated interacting systems, where many-body eigenstates
in the presence of sufficiently strong disorder can be localized 
in a region of Hilbert space even at nonzero temperature.
An MBL system comes along with universal characteristic properties 
such as area-law entanglement of highly excited states 
(HES)~\cite{Bauer:2013,Kjall:2014},
power-law decay and revival of local 
observables~\cite{Serbyn:2014_1,Serbyn:2014_2}, logarithmic growth of 
entanglement~\cite{Znidaric:2008,Bardarson:2012,Vosk:2013,Serbyn:2013_1}
as well as the violation of ``eigenstates thermalization hypothesis'' 
(ETH)~\cite{Deutsch:1991,Srednicki:1994,Rigol:2008}. 
The latter raises the appealing prospect of protecting quantum order as well as 
storing and manipulating coherent information in out-of-equilibrium many-body
states~\cite{Huse:2013,Chandran:2014,Bahri:2015,Potter:2015,Yao:2015}.

Recently it has been questioned~\cite{Carleo:2011,DeRoeck:2014_1,DeRoeck:2014_2,Hickey:2016,
Schiulaz:2014,Schiulaz:2015,Papic:2015,Yao:2016,Barbiero:2015,Smith:2017}
whether quench disorder is essential to trigger ergodicity breaking or 
one might observe glassy dynamics in translationally invariant systems, too.
In such models initial random arrangement of particles effectively fosters 
strong tendency toward self-localization characterized by 
MBL-like entanglement dynamics, exponentially slow relaxation of a typical 
initial inhomogeneity and arrival of inevitable thermalization.
This asymptotic MBL--tagged quasi-MBL~\cite{Yao:2016}--in contrast to 
the genuine ones, is not necessarily accompanied by the emergence of infinite number of conserved
quantities~\cite{Serbyn:2013_2,Huse:2014,Chandran:2015,Ros:2015}. 

Here we present a novel mechanism toward quasi-MBL in a family of \textit{clean} 
self-correcting memories, in particular the Kitaev toric 
code~\cite{Dennis:2002,Kitaev:2003} on ladder geometry, a.k.a. the Kitaev ladder 
(KL)~\cite{Karimipour:2009,Langari:2015}. The elementary excitations of KL are associated with point-like 
quasi-particles, known as electric (\textit{e}) and magnetic (\textit{m}) charges.
Our main interest has its roots in the role of non-trivial 
statistics between anyons that naturally live in (highly) excited states of such models.

Stable topological memories, by definition, need to preserve the coherence of 
encoded quantum state for macroscopic timescales. However, due to their thermal 
fragility~\cite{Castelnovo:2007,Castelnovo:2008,Brown:2016,Nussinov:2008,
Hastings:2011}, 
specially far-from-equilibrium~\cite{Kay:2009,Zeng:2016}, the problem of identifying a stable low-dimensional 
quantum memory is still unsecure.
One of the major obstacles to this end is that they do not withstand dynamic 
effects of perturbations whenever a nonzero density of anyons are 
initially present in the system. Indeed, propagation of even one pair of deconfined anyons
around non-contractible loops of the system leads to logical error. In addition, 
system's dynamics under generic perturbations could be so tangled that the quantum memory 
equilibrates in the thermal Gibbs state, in which no topological order survives~\cite{Hastings:2011}.

So far, extensive searches have been carried out to combat the mentioned shortcomings
~\cite{Chamon:2005,Kim:2016,Alicki:2010,Mazac:2012,Brown:2014,Hamma:2009,Pedrocchi:2013,Cardinal:2015,Tsomokos:2011,Wootton:2011,Stark:2011}. 
In particular, exerting an external disorder on a stabilizer code strengthens the stability of topological 
phase~\cite{Tsomokos:2011} and ensures the single particle localization of Abelian anyons
as long as their initial density is below a critical value~\cite{Wootton:2011,Stark:2011}.
The inquiry is whether one can treat the fragility of translationally invariant, 
topologically ordered systems in the presence of an arbitrary initial density 
of anyons living in HES.

We supply clear-cut evidences that random configurations of anyons in HES of KL 
prompts a self-generated disorder purely due to the mutual braiding statistics.
Performing a dual mapping suggests that nonzero density of
the magnetic charges, as barriers, poses a kinetic constraint on dynamics 
of the electric ones and hinders the propagation of the {\it{confined}} charges on 
non-trivial class of loops around the cylindrical surface of KL.
Subsequently, it is more favorable for the initial information to be encoded 
in sub-spaces with higher density of anyons!
Finally, we provide numerical evidences
that the effective disorder leads to the 
existence of an exponentially diverging time scale for dynamical persistence of the 
initial inhomogeneity, along with an intermediate slow growth of the entanglement entropy,
all of which are essential qualities of quasi-MBL.
\begin{figure}[!t]
\centering
\centerline{\includegraphics[width=1.0\linewidth]{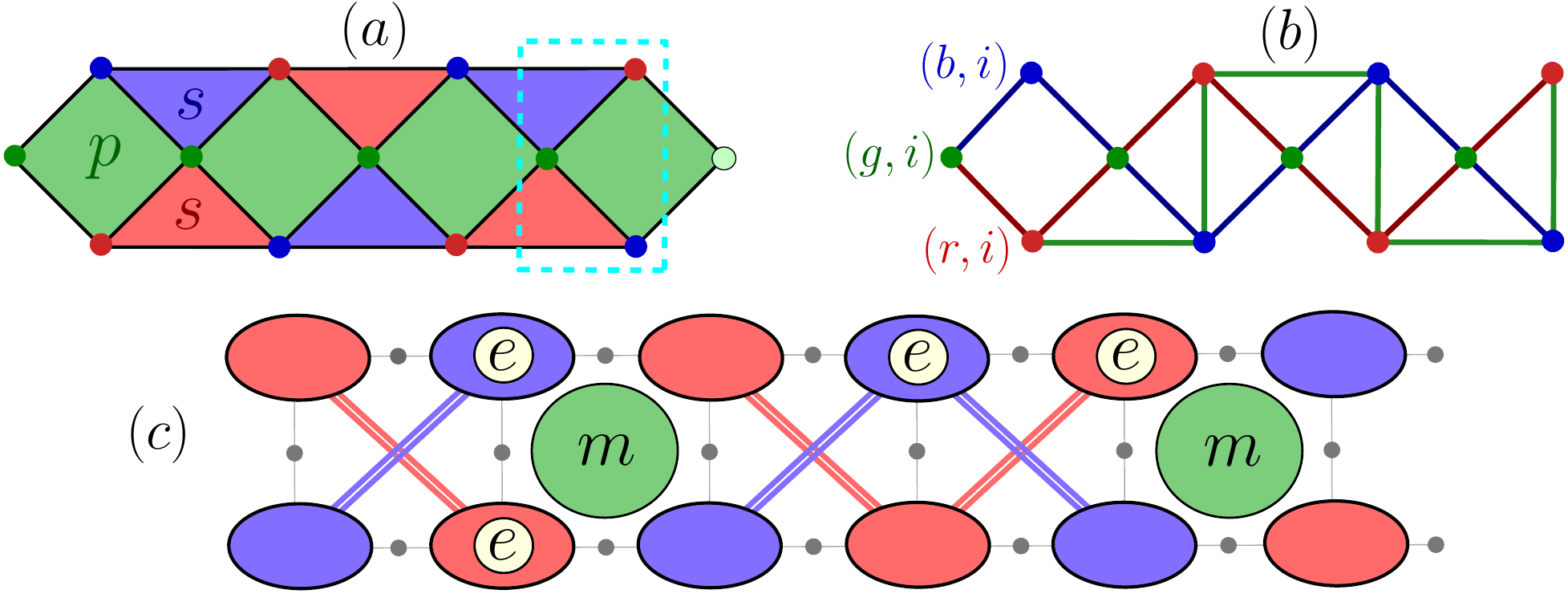}}
\caption{(a) KL with periodic-boundary condition in leg and open-boundary condition
along rungs. Spin-1/2 particles are placed on the $3L$ nodes of the lattice. (b) Multiplying 
$Z$ operators along the red (blue) line yields dual spin $\mu^x_{r/b,i}$, 
and a similar string of $X$ operators along the green line defines 
$\mu^{x}_{g,i}$. (c) The pictorial 
demonstration of the dual KL Hamiltonian, Eq.~(\ref{eq:H_IKL_dual}), for $t_m=0$.}
\label{fig:KL_02}
\end{figure}

\emph{Kitaev ladder Hamiltonian.}---KL is composed of $L$ unit cells, 
each with three sites, that we will refer to as red, green, and blue sites 
(see Fig.~\ref{fig:KL_02}-(a)), and spin-1/2 particles are
placed on the $N=3L$ nodes of the lattice. The unperturbed Hamiltonian is defined by $L$ 
plaquette stabilizer, ${B}_{p}$, and $2L$ vertex stabilizer terms on the
triangles of the ladder, $A_{s}^{r/b}$, as follow: 
\begin{align}\label{eq:H_KL}
H^{KL}_0 ~~~&=-j_m\sum_ {i} B_{p}(i)-j_e\sum_{i} 
\left(A_{s}^{r}(i)+A_{s}^{b}(i)\right),\\
B_{p}(i) ~~&=Z_{g,i}Z_{r,i}Z_{b,i}Z_{g,i+1},\cr
A_{s}^{r/b}(i)&=X_{r/b,i-1}X_{g,i}X_{b/r,i}, \nonumber 
\end{align}
where $X_{i}$ and $Z_{i}$ are the x- and z-component of Pauli operators, respectively.
We set $j_{e}$, $j_m>0$ and choose an overall energy scale by setting $j_e=1$.
KL can be viewed as the Kitaev toric code with surface termination along the rungs direction 
(a.k.a. surface code~\cite{Dennis:2002}) whose width is one. 
This model represents $\mathbb{Z}_2 \times \mathbb{Z}_2$ symmetry-protected 
topological (SPT) order associated to
anyonic parities~\cite{Langari:2015,supp_info}.

Now we would like to perturb the KL Hamiltonian such that $e$ (charge) and $m$ 
(flux), corresponding to $A_s = -1$ and $B_p = -1$, respectively, hop across the ladder and gain kinetic 
energy. To this end we consider the perturbed KL with the generic Ising terms:
\begin{eqnarray}
H^{KL} = H_0^{KL} - t_e \sum_{\langle i,j\rangle \in \partial p}Z_i Z_j - t_m \sum_{i \in legs}X_i X_{i+1},
\label{eq:H_IKL}
\end{eqnarray}
where $t_e$ ($t_m$) is the hopping strength of $e$ ($m$).
Via applying $Z_{g,i}Z_{r,i}$, which commutes with every plaquette 
and star operator except $A_s^{r}(i)$ and $A_s^r(i+1)$, an $e$ charge on site $(r,i)$ transfers to $(r,i+1)$.
We could also use $Z_{b,i}Z_{g,i+1}$ to carry out 
the same task. However, $Z_{b,i}Z_{g,i+1} = B_{p}(i)Z_{g,i}Z_{r,i}$, 
and therefore the total contribution to the Hamiltonian is 
$(1+B_p(i))Z_{g,i}Z_{r,i}$. That being so, the Ising interactions are coupled 
to the dynamical $\mathbb{Z}_2$ {\it{gauge field}} and the hopping of an $e$ charge from 
$(r,i)$ to $(r,i+1)$ depends on the value of $B_p(i) =(-1)^{n^m_i}$, which takes into account the parity of 
$m$ anyon on site $(g,i)$, or equivalently, the mutual braiding statistics of $e$ and $m$.
Hence, the hopping of $e$ is blocked if there is a flux on its way. 
Likewise, $X_{r,i}X_{b,i+1}$ transports one unit of $m$ charge from plaquette 
$(g,i)$ to $(g,i+1)$ and vice versa. We could also consider $X_{b,i}X_{r,i+1}$ 
to reach the same goal. However, $X_{b,i}X_{r,i+1} = A_s^r(i+1)A_s^b(i+1)X_{r,i}X_{b,i+1}$, and 
again the movement of $m$ is intertwined with the density of $e$'s along its way. 

Unlike the single particle studies in the presence of disorder~\cite{Wootton:2011,Stark:2011},
in a many-body picture, the transport properties of Abelian anyons might 
strongly be affected by their exotic statistics, so that $e$ and $m$ as two distinct quasi-particles are able to mutually 
suppress their own dynamics, even in a clean system. The latter property is the characteristic feature of Falicov-Kimball 
like Hamiltonians whose non-ergodic dynamics has been recently investigated as a candidate for disorder-free 
localization~\cite{Schiulaz:2014,Schiulaz:2015,Papic:2015,Yao:2016,
Barbiero:2015,Smith:2017}.

To directly reveal this hidden structure, we introduce a non-local dual 
transformation, which maps Eq.~\ref{eq:H_IKL} to three coupled transverse 
field Ising (TFI) chains (see Fig.~\ref{fig:KL_02}-(b)),
\begin{eqnarray}\label{eq:H_IKL_dual}
H^{KL}_{\rm dual} ~= ~&&- \sum_{i} \left(j_m \mu_{g,i}^{z} + t_m 
\mu_{g,i}^x\mu_{g,i+1}^x(1 + \mu_{r,i+1}^{z}\mu_{b,i+1}^z)\right)\cr
&& -\sum_{i}\left(j_e \mu_{b,i}^{z} + t_e \mu_{b,i}^x\mu_{b,i+1}^x(1 + 
\mu_{g,i}^{z}) \right)\cr
&& -\sum_{i}\left(j_e \mu_{r,i}^{z} + t_e \mu_{r,i}^x\mu_{r,i+1}^x(1 + 
\mu_{g,i}^{z})\right),
\end{eqnarray}
where 
\begin{eqnarray}
\mu_{g,i}^x = \prod_{j \ge i} X_{r,j}X_{b,j+1},\quad \mu_{r/b,i}^x = \prod_{j 
\ge i} Z_{g,j}Z_{r/b,j}, \nonumber
\end{eqnarray}
and $\mu_{g,i}^z = B_p(i)$, $\mu_{r/b,i}^z = A_s^{r/b}(i)$. This model features glassy dynamics for 
the $\mu_{g,r/b}$ degrees of freedom, associated to the anyonic excitations of the KL.
For example, in the non-interacting limit $t_m=0$ an effective description of $H^{KL}_{\rm dual}$ 
reduces to two TFI chains coupled to the static $\mathbb{Z}_2$ gauge field, $\mu^{z}_{g}$. 
These gauge degrees of freedom form a set of $L$ constants of motion with trivial 
dynamics. Hence, an arbitrary initial nonzero density of fluxes $\rho_m$ 
\textit{energetically} suppresses charges' kinetic interactions (see~Fig.\ref{fig:KL_02}-(c)). Indeed,
a typical initial inhomogeneity of $\{\mu_{g}\}=\pm1$ is dynamically manifested in 
a \textit{self-generated disorder potential}, $t'_{e}(i)=t_{e}(1+\mu_{g,i})$, 
with the dilution distribution,
\begin{align}
\mathcal{P}(t'_e)=(1-\rho_{m})\delta(t'_e-2t_e) + \rho_{m}\delta(t'_e),
\label{eq:dilution}
\end{align}
where for a fixed value of $\rho_{m}$, different configurations of 
fluxes correspond to different realizations of disorder. In such situation,
dynamics of the whole system will be identical to that of two decoupled,
disordered TFI chains in terms of $\hat{\mu}_{r/b}$, which are Anderson localized~\cite{Stinchcombe:1981}. 
\begin{figure}[t!]
\centering
\includegraphics[width=1.0\linewidth]{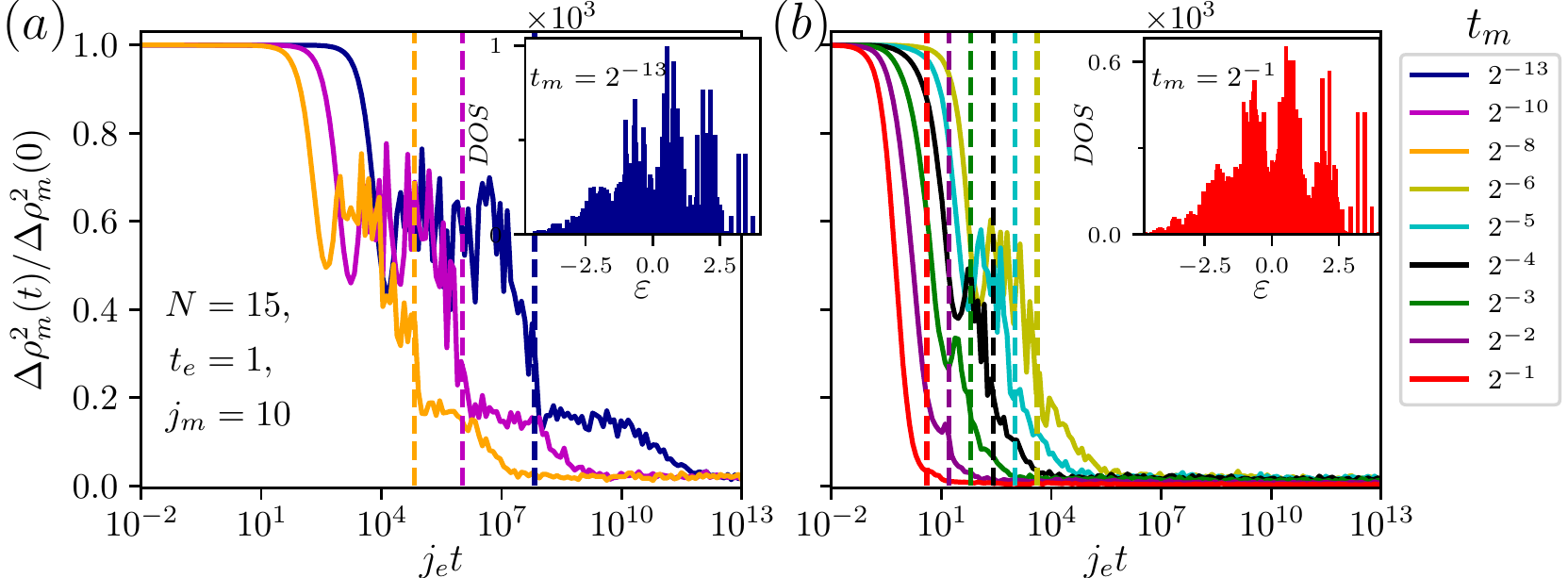}
\caption{Relaxation of the flux inhomogeneity density 
for $N=15$ (spins) with 
fixed $t_{e}=1$ and varying $t_m$, averaged over $150$ 
random initial states $\lvert \psi_{N_m} \rangle$, with $N_m=3$ 
in the (a) fast and (b) slow dynamics regime.
Vertical dashed lines represent time at $\tau_{eff}$.
(insets) DOS for the cases with largest and smallest mass ratio $t_m/t_e$.
}
\label{fig:inhomo_density} 
\end{figure}

\emph{ Anyonic self-localization and quasi-MBL regime.}---The outlined glassy blueprint is inspiring to look 
for the counterpart of quasi-MBL~\cite{Yao:2016} in an Abelian many-body 
system. In analogy to the observed quasi-MBL in a trivial admixture of heavy and 
light particles~\cite{Schiulaz:2014,Schiulaz:2015,Papic:2015,Yao:2016,Barbiero:2015},
one needs to choose the mass ratio of the quasi-particles to be large enough, as 
long as the ``isolated bands''~\cite{Papic:2015} due to finite-size effects is 
not manifested. In the KL, $t_e$ ($t_m$) controls the strength of 
the effective disorder (interaction) as well as the effective mass of 
$e$ ($m$) anyons~\cite{comment}. Thus, we consider the limit {\text{\small{$0 
< t_m \ll t_e$}}}, where $m$'s have a large but finite effective mass.
Now we initialize the whole system in a typical inhomogeneous  configuration 
$\lvert \psi_{N_m} \rangle$ of $N_m=\rho_mL$ fluxes, which are selected near 
the middle of the spectrum of $H^{KL}$.
Then, we compute the evolution of the flux inhomogeneity density under 
$H^{KL}$, 
\begin{eqnarray}
\Delta \rho_{m}^{2}(t) \equiv \dfrac{1}{L}\sum ^{L}_{p=1}\lvert \langle 
\psi_{N_m}
\rvert ( n^{m}_{p+1}(t)-n^{m}_{p}(t) ) \lvert \psi_{N_m} \rangle \rvert ^{2},
\label{eq:den_inhomo}
\end{eqnarray}
which vanishes for any perfect delocalized state. 
As illustrated in Fig.~\ref{fig:inhomo_density}-(a,b), for {\text{\small{$t_m\gtrsim0.01$}}} fast 
relaxation of the initial inhomogeneity due to resonance 
admixtures~\cite{Schiulaz:2014,Schiulaz:2015} takes place within the 
time scale {\text{\small{$\tau_{int} \sim t_m^{-1}$}}}, while in the opposite limit, i.e. small $t_m$,
the initial inhomogeneity plateau persists until $\tau_{int}$.
Moreover, the residual inhomogeneity remains even at later times 
{\text{\small{$\tau_{eff}\sim t_{e}/t_{m}^2$}}}. 
Subsequent to this time, the collective slow dynamics eventually
gives way to complete relaxation at $\tau_R$. As discussed in~\cite{Papic:2015},
to ensure the robustness of the numerical results against finite-size effects, the density of states (DOS) must not display
any isolated classical bands, which can be clearly seen in the insets of Fig.~\ref{fig:inhomo_density}.

\begin{figure}[!t]
\centering
\includegraphics[width=0.9\linewidth]{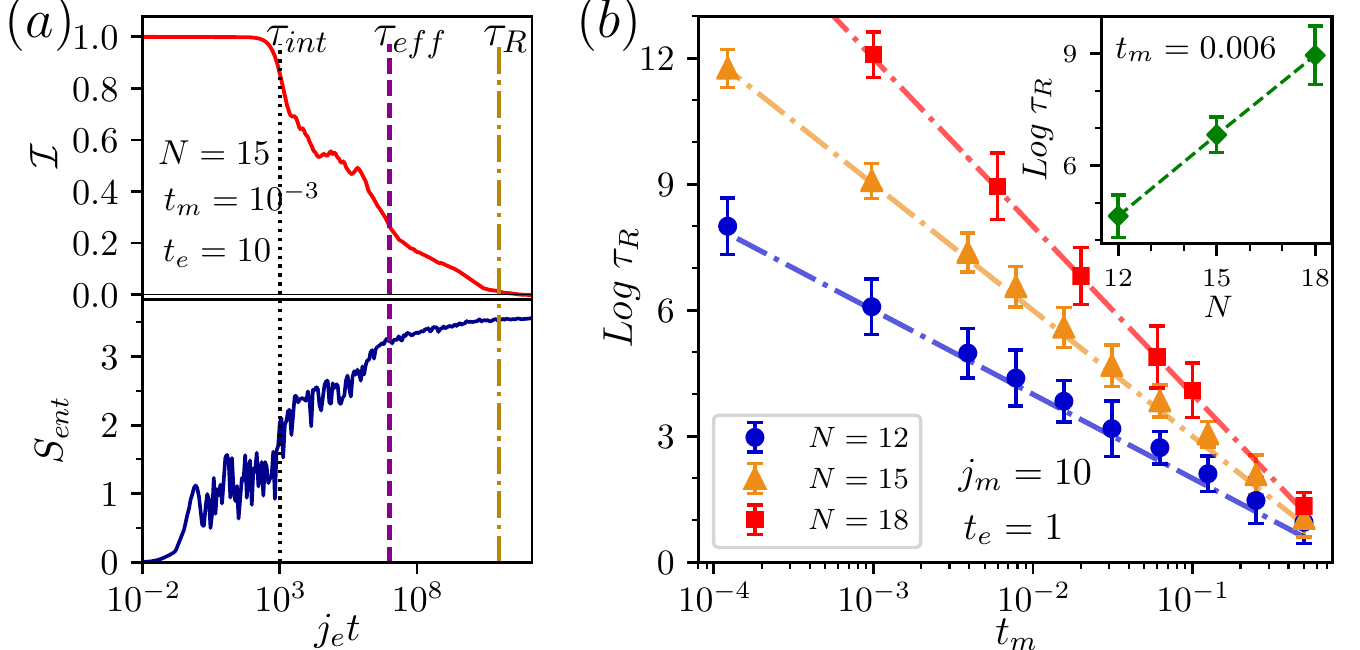}
\caption{(a) Upper panel: Relaxation of the time averaged inhomogeneity of fluxes for $N=15$ and $t_{e}=10$, 
in quasi-MBL regime at $t_m=10^{-3}$, averaged over $150$ initial states with $N_m=3$. Lower panel: entanglement 
dynamics for subsystem of size $N_A=8$, averaged over $250$ random
product states. (b) The scaling behavior of $\tau_R$ versus $t_m$ at fixed $t_e=1$, for $N=12,15,18$ with the initial
total number of fluxes, $N_m=2,3,4$, respectively. The dot-dashed lines denote the analytical estimation
${\text{\small{$\tau_R t_m\propto(t_e/t_m)^{N_m-1}$}}}$. (inset) The expected exponentially diverging behavior
of $\tau_R$ with system sizes at $t_m=0.006$ in the quasi-MBL regime.
}
   \label{fig:Sent_bear_TR_scaling}
\end{figure}

To gain further insights on the nature of the three distinct time scales that 
characterize the relaxation dynamics of anyons, {\text{\small{$\tau_{int} < \tau_{eff} \leq \tau_R$}}}, 
we have also looked at the growth of entanglement entropy $S_{ent}=-\rho_A \ln\rho_A$ for half-system $A$
with strong disorder $t_e = 10$ (see Fig.~\ref{fig:Sent_bear_TR_scaling}-(a)). 
Prior to $\tau_{int}$, charges perceive the fluxes as if they 
are immobile barriers. 
Hence, after an initial growth, $S_{ent}$ saturates to the 
first plateau, conveying the single particle localization length of charges.
Subsequent to $\tau_{int}$ hybridization of fluxes arrives, which in turn intertwines with the charge dynamics. 
Thus, the entropy shows logarithmic growth until the finite-size dephasing of 
charges wins at the second plateau. At $\tau_{eff}$, the dephasing of the fluxes sets in and the entanglement 
grows even more slowly to saturate eventually at $\tau_R$.

It is worth mentioning that the same time scales also determine the evolution of the quantity
$\mathcal{I}(\tau)\!=${\text{\tiny{$\frac{\overline{\Delta 
\rho_{m}^{2}(\tau)} - \overline{\Delta 
\rho_{m}^{2}(\infty)}}{\overline{\Delta 
\rho_{m}^{2}(0)} - \overline{\Delta \rho_{m}^{2}(\infty)}}$}}}, 
where 
\begin{small}$\overline{\Delta \rho_{m}^{2}(\tau)}\equiv {\tau^{-1}}\int_{0}^{\tau} dt~{\Delta \rho_{m}^{2}(t)}$\end{small} 
(see Fig.~\ref{fig:Sent_bear_TR_scaling}-(a)). Moreover, the late time behavior of $\mathcal{I}(\tau)$ 
signifies that the full relaxation eventually occurs but at very long times. 
This anyonic slow dynamics is in contrast to true MBL in which an initial inhomogeneity never 
completely decays. These results resemble those characteristic behaviors 
observed in the previous proposal of quasi-MBL~\cite{Yao:2016}.

Following the perturbative argument presented in~\cite{Schiulaz:2014}, the 
final decay time of initial inhomogeneity should display
scaling behavior as {\text{\small{$\tau_R t_m\propto(t_e/t_m)^{N_m-1}$}}} in KL. To confirm this parametric dependence, 
we plot numerically extracted value of $\tau_R$ versus $t_m$ in Fig.~\ref{fig:inhomo_density}-(b)~\cite{supp_info}
for different system sizes at fixed value of $t_e = 1$.
Our numerical results not only illustrate a good agreement with 
the rough estimation presented above, but also imply the exponential dependence of $\tau_R$
on the system size with growing number of fluxes in the quasi-MBL regime.

Lastly, we would like to address whether the fast relaxation in such 
anyonic liquids is followed by the viability of ETH. 
In this respect, we discard the pair creation/annihilation of fluxes, which 
might happen as a result of {\text{\small{$X_{r,i} X_{b,i+1}$}}} terms in 
Eq.~(\ref{eq:H_IKL}). 
Typically, this recombination processes are less likely for the heavy 
particles in comparison with the light ones. Therefore, it is a reasonable 
assumption to consider {\text{\small{$\tilde{H}_{KL}=\hat{P}_{N_{m}} 
H^{KL}\hat{P}_{N_{m}}$}}},
where {\text{\small{$\hat{P}_{N_{m}}$}}} is the projection operator into the 
subspace with fixed $N_m$~\cite{supp_info}. 
\begin{figure}[t!]
\centering
\includegraphics[width=0.745\linewidth]{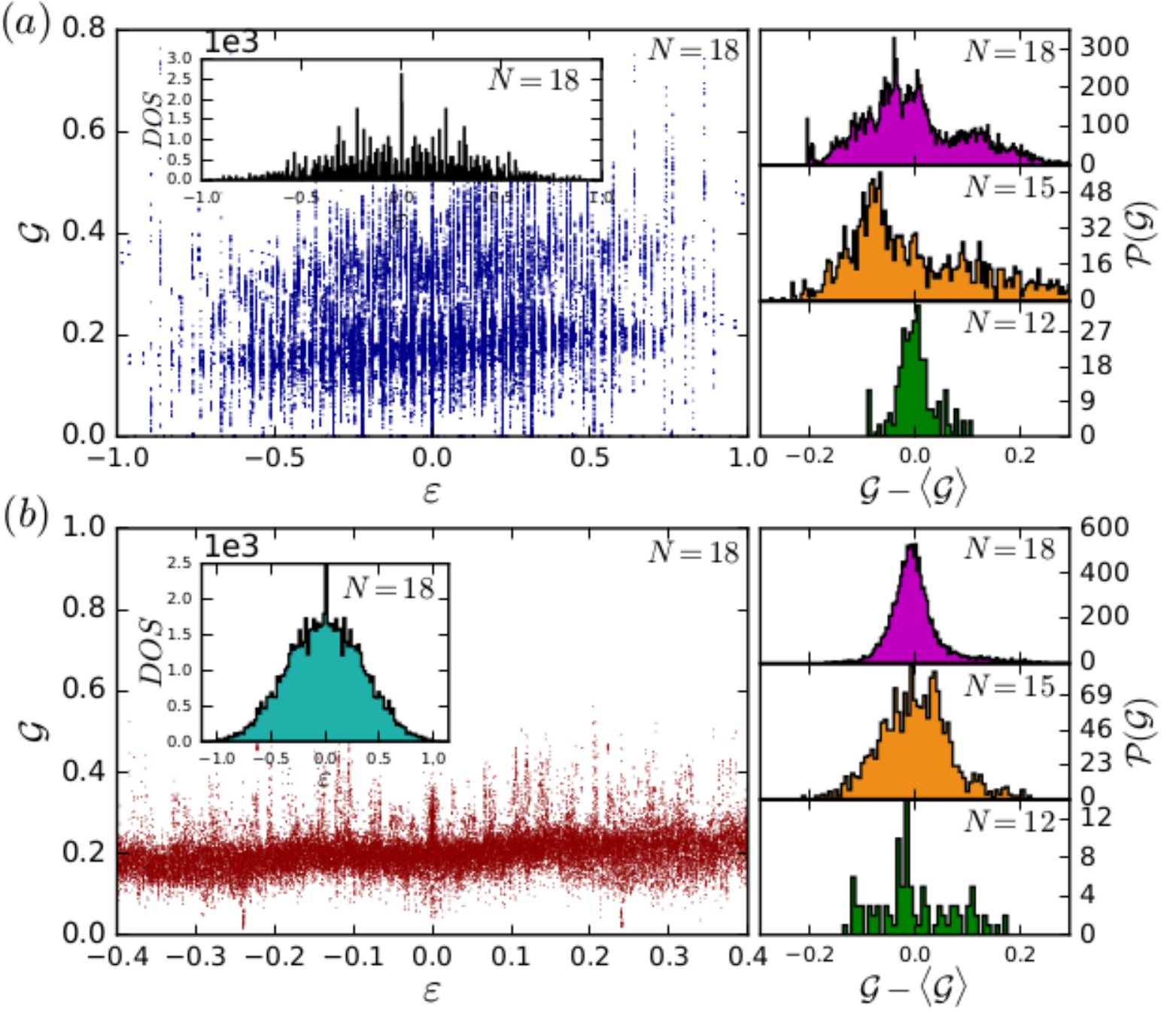}
\caption{(a) Failing of ETH for the case 
{\text{\small{$(t_{e},t_{m})=(1,0.001)$}}}. 
Left panel: The exception value of {\text{\small{$\mathcal{G}$}}} in momentum 
eigenstates versus 
energy density. Inset displays DOS for 
{\text{\small{$\tilde{H}_{KL}$}}}. 
Right panel: {\text{\small{$\mathcal{P}(\mathcal{G})$}}} for all eigenstates in 
the energy window 
{\text{\small{$\varepsilon \in [-0.09,0.09]$}}} in the middle of the energy band 
for different system sizes. The average value is removed from the distribution for 
visibility. (b) Manifestation of ETH for {\text{\small{$(t_{e},t_{m})=(1,0.5)$}}}. 
{\text{\small{$\mathcal{P}(\mathcal{G})$}}} is 
plotted in the interval {\text{\small{$\varepsilon \in (0,0.15]$}}}. Results are 
reported for {\text{\small{$(N,N_{m})=(12,2)$}}}, 
{\text{\small{$(15,3)$}}} and {\text{\small{$(18,3)$}}}.}
   \label{fig:ETHab}
\end{figure}  
\begin{figure*}[t!]
\centering
\centerline{\includegraphics[width=0.94\linewidth]{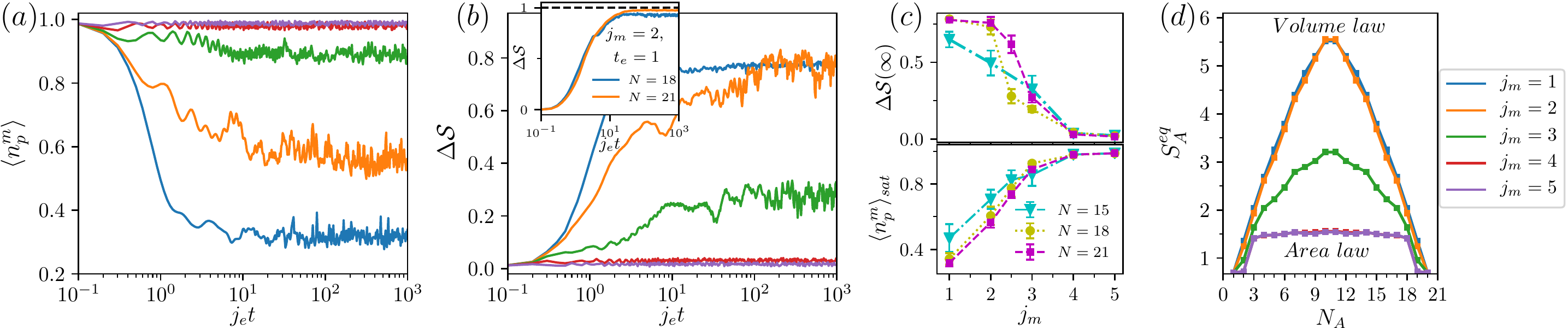}}
\caption{(a) Evolution of the expectation value of $\hat{n}^{m}_{p}$ 
by varying the flux mass gap $j_m$, 
for $N=21$, $t_e=10$, $t_m=0.6$ and the pre-quench state initialized in $\rho_m=1$ and $\rho_e=0$. 
Since both the Hamiltonian and initial state are translationally invariant, the results for all 
plaquettes are the same.
(b) The information spreading, (c) late time saturation values and (d) scaling of $S_{ent}$ with different 
subsystem sizes $N_A$, at $j_et=10^3$, where $\langle n_{p}^{m}\rangle$ 
saturates to its long-time value. (inset) The heating procedure
in the ergodic phase, in which systems with sizes $N = 18$ and $21$ could 
approach their thermal Gibbs state.
}
\label{fig:KrylovJZ10V2}
\end{figure*}

We evaluate {\text{\small{$\mathcal{G} \equiv \langle n_{p}^{m} n_{p+1}^{m}\rangle$}}} in the 
simultaneous eigenstates of {\text{\small{$\tilde{H}_{KL}$}}} and 
momentum, and collect the results from all momenta. Fig.~\ref{fig:ETHab}-(a) 
quantifies {\text{\small{$\mathcal{G}$}}} in different energy densities, for 
{\text{\small{$(t_{e},t_{m})=(1,0.001)$}}}.
In the whole energy intervals, the value of {\text{\small{$\mathcal{G}$}}} is spread considerably in a 
wide range. On top of that, the distribution function {\text{\small{$\mathcal{P}(\mathcal{G})$}}} near the middle of
the band has a very broad half-width and peaks at different values as the system size is increased 
(right panel in Fig.~\ref{fig:ETHab}-(a)) indicating a strong non-ergodic behavior.
For the fast dynamics, e.g. {\text{\small{$(t_{e},t_{m})=(1,0.5)$}}},
the system rather obeys ETH prediction: DOS becomes continuous,
{\text{\small{$\mathcal{P}(\mathcal{G})$}}} sharply picked around its mean value 
and the width of the distribution decreases by increasing the system size, as 
illustrated in Fig.~\ref{fig:ETHab}-(b).
Hence, whenever $t_e$ and $t_m$ are in the same order of magnitude,
the fast relaxation occurs along with the validity of ETH in HES.

\emph{Resilience of the topological order following a quench}.---To further inspect the 
fingerprints of the emergent kinetic constraint in Eq.~\ref{eq:H_IKL_dual}
on non-equilibrium anyonic dynamics, we proceed with the scenario of the quantum quench. 
We initially prepare the system in 
$\lvert \phi_0 \rangle = (1/2)~(1+\hat{W}_x)\lvert\{n_{p}^{m}\},\{n_{s}^{e}\} \rangle$, 
where $\hat{W}_x$ is one of the two logical operators that encode the 
topological qubit, and $\lvert\{n_{p}^{m}\},\{n_{s}^{e}\} \rangle$ is an
exact eigenvector of $H^{KL}_{0}$ including a specific pattern of $e$'s and $m$'s 
(with $\rho_m$ ($\rho_e$) density of $m$ ($e$) preparatory anyons). 
According to the SPT nature of $H^{KL}_{0}$~\cite{supp_info},
the corresponding eigenstates are short-range entangled, and thus cold state.
We are specially interested in those transitionally invariant pre-quench 
states with $\rho_m=1$ and $\rho_e=0$, wherein initial information is encoded in 
the sub-space with maximum number of $m$ anyons. By performing massively parallel 
time integration based on Chebyshev expansion~\cite{petsc-user-ref,petsc-efficient,Hernandez:2005} 
to evolve system under $H^{KL}$, we measure the spreading of the stored initial information
as well as the non-equilibrium heating procedure over the course of time,
\begin{eqnarray}
\Delta \mathcal{S}= \dfrac{S_{ent}(t)-S_0}{S_{page}-S_0},
\end{eqnarray}
where $S_{ent}(t)$, $S_0$ and $S_{page}$ are 
bipartite entanglement entropies associated with $\lvert \phi(t) \rangle$,
$\lvert \phi_0 \rangle$ and an infinite temperature state~\cite{Page:1993}, respectively.

For $t_m=0$, it follows from Eq.~\ref{eq:dilution} that the dynamical 
effect of perturbation would be completely suppressed due to the presence of 
fluxes in each plaquette.
Indeed, regardless of the magnitude of $j_m$ and $t_e$, the spreading of information is strictly impeded, even 
in the absence of any explicit form of disorder in either the Hamiltonian or the
initial state! This result is comparable to Ref.~\onlinecite{Smith:2017}, where 
non-ergodic dynamics is induced solely through the presence of 
static gauge degrees of freedom, albeit through a distinct mechanism.

For $t_m\neq0$ thermalization process can be controlled by the 
flux mass gap $j_m$~\cite{Dennis:2002}. 
In this respect, to build up the tendency toward the survival of preparatory 
fluxes in the system we increase $j_m$ (see Figs.~\ref{fig:KrylovJZ10V2}-(a,c)).
As a result, the thermalization process slows down
at a continually growing rate (see Figs.~\ref{fig:KrylovJZ10V2}-(b,c)). 
Hence, scaling of $S_{ent}$ with different subsystem sizes $N_A$ reforms from 
volume law to area law, as depicted in Fig.~\ref{fig:KrylovJZ10V2}-(d). 
An approximate area law is observed for $j_m \gtrsim 4$, while
for $j_m=1$, $2$ the scaling obeys volume law~\cite{footnote}. Notably, the system with sizes 
considered here, in the thermal regime, reaches its infinite temperature state (see inset of Fig.~\ref{fig:KrylovJZ10V2}-(b)),
which is never observed in those suffering from finite-size effects.

\emph{Discussion.}---We identified the anyonic self-localization as an emergent property purely rooted in the 
intrinsic statistics of the Abelian anyons and extended the so-called quasi-MBL picture to self-correcting 
quantum codes. This provides a novel mechanism distinct by nature
from the recent proposals on disorder-free localization
~\cite{Schiulaz:2014,Schiulaz:2015,Yao:2016,Grover:2014,Smith:2017}.
 
As mentioned earlier, a number of approaches have been proposed to remedy 
the thermal fragility of the quantum memories with $D<4$, such as: considering clean 
cubic~\cite{Chamon:2005,Kim:2016} and higher dimensional 
codes~\cite{Alicki:2010,Mazac:2012}
with a more complex structure than the toric one, 2D codes
consisting of N-level spins~\cite{Brown:2014}, coupling  2D
codes to a massless scalar field~\cite{Hamma:2009,Pedrocchi:2013,Cardinal:2015} as well as 
employing Anderson localization machinery~\cite{Tsomokos:2011,Wootton:2011,Stark:2011}. 
Our results  on anyonic self-induced disorder indicate that an exponentially glassy dynamics 
could be induced even ({\it{i}}) in a {\it{low-dimensional}}, 
{\it{clean}}, and {\it{simple-structure}} models, and remarkably 
({\it{ii}}) this glassiness is enhanced by {\it{increasing}} 
either the density of errors and/or environmental perturbations~\cite{Schiulaz:2015,supp_info}.

The scenario discussed in this work could be easily generalized~\cite{supp_info} to 2D quantum double models 
such as the Levin-Wen model~\cite{Levin:2005}. 
As little progress has been made toward investigation of MBL in topologically
ordered 2D systems, our study breaks the ground for future researches.

A closely related concept to quasi-MBL in multi-component systems 
is quantum disentangled liquid~\cite{Grover:2014,Garrison:2016,Veness:2016},
wherein ``post-measurement'' of the anyon configuration is identical to the 
error syndrome; that is, the first step in the error-correcting protocol. It is tempting to see whether 
such measurement procedure on a topological state supports this claim. 
Intuitively, could there a quantum disentangled \textit{spin} liquid be found?

\emph{Acknowledgements.}---We highly appreciate fruitful discussions and neat 
comments by Fabien Alet, Moeen Najafi-Ivaki, Mazdak Mohseni-Rajaee
and Arijeet Pal. The authors would like to 
thank Sharif University of Technology for financial supports
and CPU time from Cosmo cluster.
AV was supported by the Gordon and Betty Moore Foundation's EPiQS Initiative 
through Grant GBMF4302.

%
\pagebreak
\onecolumngrid
\newpage
\renewcommand{\thesection}{S\arabic{section}}    
\renewcommand{\thefigure}{S\arabic{figure}}
\renewcommand{\theequation}{S\arabic{equation}} 
\setcounter{figure}{0}
\setcounter{equation}{0}
\begin{center}
{\large \bf Supplemental Material for EPAPS\\ 
Anyonic self-induced disorder in a stabilizer code:\\
quasi-many body localization in a translational invariant model}\\
\vspace{0.4cm}

H. Yarloo,$^1$ A. Langari,$^{1,2,*}$ and A. Vaezi$^3$

{\small \it $^1$Department of Physics, Sharif University of Technology, P.O.Box 
11155-9161, Tehran, Iran}

{\small \it $^2$Center of excellence in Complex Systems and Condensed Matter 
(CSCM), Sharif University of Technology, Tehran 14588-89694, Iran}

{\small \it $^3$Department of Physics, Stanford University, Stanford, CA 94305, 
USA}
\end{center}
\vspace{0.4cm}
\noindent  {\bf Kitaev Ladder (KL).}---We first briefly discuss the
main properties of the KL Hamiltonian, $H_0^{KL}$, in 
(highly) excited states to give an insight on its nature as an anyonic liquid.
In KL periodic-boundary condition in the leg direction leads to the global 
constraint $\prod_{i} A_{s}^{r/b}(i)= \mathbb{1}$, which ensures that (\textit{i}) 
any energy level in the whole spectrum will have at least twofold
degeneracy and (\textit{ii}) $2L-1$ independent charge degrees of freedom are
created/annihilated in pairs. However, fluxes can be created/annihilated 
singly in contrast to the 2D toric code on torus. One can also define the
occupation operators of charges and fluxes as 
$\hat{n}_{i}^{e}=(1-A_s^{r/b}(i))/2$ and $\hat{n}_{i}^{m}=(1-B_{p}(i))/2$, 
respectively. In term of occupation operators, $H_0^{KL}$ simply counts the 
total number of \textit{e} and \textit{m} anyons in the system. 
The basis which diagonalizes $H_0^{KL}$ has the following closed form in the
occupation number representation:
\begin{eqnarray}
\lvert\{n_{i}^{m}\},\{r_{i}\}\rangle = \hat{P}_{\{n_{i}^{m}\}} 
\hat{g}_{\{r_{i}\}}\lvert x+\rangle^{\otimes N}, \quad 
\label{eq:close_form_ev}
\end{eqnarray}
where 
$\hat{P}_{\{n_{i}^{m}\}} = \prod_{i}( 1+(-1)^{n_{i}^{m}}B_{p}(i))/2$, 
$\hat{g}_{\{r_{i}\}} = \prod_{\substack{i\in \text{D}}}Z_{i}^{r_{i}}$
and $\lvert x+\rangle$ is the eigenstate of $X$, i.e. $X \lvert x+\rangle= 
\lvert x+\rangle$.
$\hat{g}_{\{r_{i}\}}$ is a member of the Abelian group $G$ (consisting of 
$2^{2L}$ 
different set of configurations, $\{r_{i}=0,1\}$), which performs all spin-flip 
operations for $2L$ spins placed on the global path $D$ (see Fig.~\ref{fig:KL_lattice}).
By the action of $G$ on the charge vacuum state, one can generate all 
$2^{2L-1}$ different independent charge configurations. 
$\hat{P}_{\{n_{i}^{m}\}}$ is the projection operator to the subspace with 
flux configuration, $\{n_{i}^{m}\}$,
where $2^{L}$ different set of $\{n_{i}^{m}=0,1\}$ 
can generate corresponding flux degrees of freedom. However,
$\{r_{i}=1 ;\;\forall i\}$ is a special case for which 
$ W_{z} \equiv \hat{g}_{\{r_{i}=1\}} $ 
goes over the whole system on homologically nontrivial path $D$. 
Accordingly, for any state $\rvert \{n_{i}^{m}\rbrace,\{r_{i}\rbrace \rangle$, 
there is a degenerate state,
\begin{eqnarray}\label{eq:two_fold_deg}
\lvert\{n_{i}^{m}\} , \{\overline{r}_{i}\} \rangle =\hat{W}_{z} 
|\{n_{i}^{m}\},\{r_{i}\}\rangle,
\quad \hat{W}_{z}=\prod_{\substack{i \in \text{$D$}}}Z_{i},
\end{eqnarray}
where $\hat{W}_{z}$ plays the role of 
\textit{Wilson loop}, which creates a pair of charges on a vertex, rounds 
them across the system, and then annihilates them. This process is accompanied
by changing topological sector of states (associated to different eigenstates of 
logical operator $\hat{W}_{x}=\prod_{\substack{i \in \text{$D^\prime$}}}X_{i}$) and can be interpreted as
diffusion of unchecked errors across the ladder after a 
time proportional to the system size, a.k.a. logical error. 
Hence, generating glassy dynamics in highly-excited states (HES) of such 
disorder-free stabilizer codes impedes the mentioned procedure and puts forward a new paradigm for
realizing more stable quantum memories at finite temperature,
the task which assigned to the quasi many-body localization (qMBL) mechanism 
through this work. 
\begin{figure}[ht]
\centering
\centerline{\includegraphics[width=3.1in]{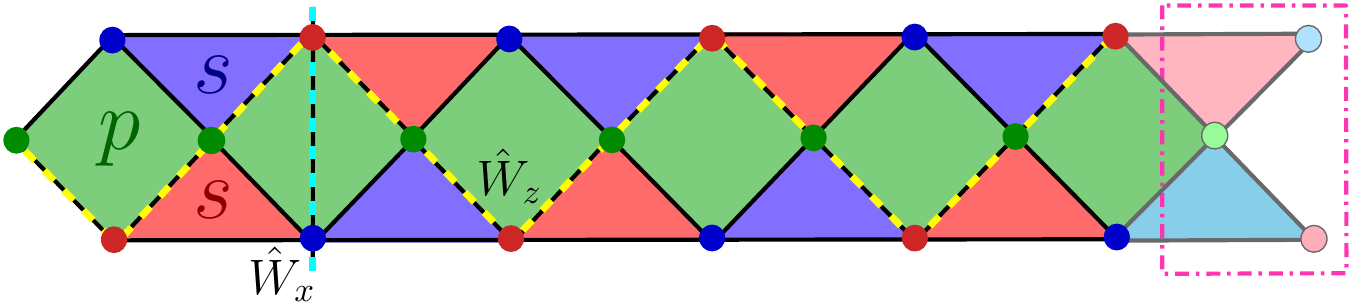}}
\caption{KL as a surface code terminated in one direction. 
Dashed lines indicate topologically nontrivial path $D$ and $D^\prime$.
}
\label{fig:KL_lattice}
\end{figure}
as mentioned in the main text, the ground state 
of KL model (with twofold degeneracy in the free charge and flux 
sectors $\{ n_{i}^{m}=0,r_{i}=0;\; \forall~i \}$) can not be smoothly 
connected to a short range entangled state without breaking the 
following Ising symmetries, explicitly or spontaneously:
\begin{eqnarray}
\mathcal{P}_m = \prod_{i=1}^{L}B_{p}(i) \label{eq:SPT_symmetries}~,\quad 
\mathcal{P}_e = \prod_{s} {A_{s}^{r}(i)} = \prod_{s} {A_{s}^{b}(i)}.
\end{eqnarray}

The action of these symmetries on occupation number basis 
Eq.~(\ref{eq:close_form_ev}) 
can be interpreted as anyonic parity, such that $\mathcal{P}_m$ shows the 
\textit{flux 
parity} and $\mathcal{P}_e$ represents the \textit{charge parity} in the red 
or blue vertices. 
%
%

\noindent  {\bf Boundary condition for the dual pseudo-spins.}---In this 
section, we study the effect of the flux parity, defined in
Eq.~(\ref{eq:SPT_symmetries}), on determining the boundary condition (BC) 
for pseudo-spins $\mu_{r/b}$. Because of the global constraint 
$\prod _{i}\mu _{r,i}^{z}\mu _{b,i}^{z}=\mathbb{1}$, $\mu_{r/b}$'s 
describe only $2L-1$ independent degrees of freedom. That being so, applying dual 
transformation on the original spins of the KL 
with periodic-BC results a 1-to-2 mapping between $\mu_{r/b}$'s and 
original spins. One can consider 
two additional independent ancillary degrees of freedom, in the virtual 
$(L+1)$-th unit cell of the ladder (see Fig.~\ref{fig:KL_lattice}). 
According to our definition for blue and red pseudo-spins,
$\mu_{r/b,i}^x = \prod_{j \ge i} Z_{g,j}Z_{r/b,j}$, the x-component of 
ancillary pseudo-spins are product over an empty set, hence $\mu _{r,L+1}^{x}=\mu_{b,L+1}^{x}=1$. On 
the other hand, the string operators $\mu _{r,1}^{x}$ and $\mu _{b,1}^{x}$ 
are two special cases that commute with $H^{KL}$ and in terms 
of the original spins take the form $\mu _{r,1}^{x}=W_{z}$, 
$\mu_{b,1}^{x}=\mathcal{P}_m W_{z}$.
Combination of these properties gives:
\begin{equation}
\label{eq:BC}
1=\mu _{r,L+1}^{x} = \left\{ \begin{array}{ll}
         \mu _{r,1}^{x} & \mbox{if $\; \; W_{z} = +1$}\\
        -\mu _{r,1}^{x} & \mbox{if $\; \; W_{z} =-1$}.\end{array} \right. 
\end{equation}

In fact, $\mu _{r,1}^{x}$ and $\mu _{b,1}^{x}$ are \textit{dynamical 
variables} that are not independent from each other and determine the BCs on 
the $\mu$'s. So, according to Eq.~(\ref{eq:BC}), the flux parity 
\begin{align}
   \mathcal{P}_m = \mu _{r,1}^{x}\mu _{b,1}^{x}\equiv (-1)^{N_{m}}=\left \{ 
   \begin{array}{ll}
    +1 \\ -1, 
   \end{array} 
   \right.
\label{eq:PBC_APBC}
\end{align}
determines the BC of the red and blue TFI chains in $H^{KL}_{\rm dual}$
and relates them to each other in such a way 
that for the even parity two chains simultaneously have periodic-BC or 
antiperiodic-BC, but for the odd parity one chain has periodic-BC while another 
must have antiperiodic-BC and vice versa.
In fact, the mentioned BCs, by doubling the size of the Hilbert space,
establish a one-to-one mapping between all energy levels of $H^{KL}_{\rm dual}$ 
and the perturbed KL Hamiltonian.

\noindent  {\bf{Numerical methods for extracting the complete relaxation time}}.---The 
numerical data presented in Fig.~\ref{fig:Sent_bear_TR_scaling}-(b) of 
the main text, have been extracted as follow:
For $N=12$ and $15$, we performed extensive exact diagonalization and averaged over $150$ 
random initial configurations of magnetic charges with fixed total 
number of fluxes, $N_m$. For $N=18$ --due to lack of special symmetries, e.g. 
$U(1)$, that can significantly decrease the effective Hilbert space
dimension--, we implemented another advanced methods:
(\textit{i}) In the fast relaxation regime ($t_m\sim t_e$), we use massively parallel 
time integration method~\cite{petsc-user-ref,petsc-efficient,Hernandez:2005} based 
 on Chebyshev expansion that is suitable for simulating the dynamics up to the moderate times ($j_e t \sim 10^4$) that are
 much larger than the typical value of $\tau_R$ in this regime.
(\textit{ii}) In order to compute $\tau_R$ for slow relaxation regime we simulate relaxation 
 time trace via shift-and-invert Lanczos method, which 
 has been employed in the previous study of quasi-MBL~\cite{Yao:2016}.
 To find a large portion of spectrum, we use the implementation of 
 PETSc~\cite{petsc-user-ref,petsc-efficient} and SLEPc~\cite{Hernandez:2005} rely 
 on MUMPS~\cite{MUMPS:2006} to perform parallel sparse Cholesky factorization as a direct solver. 
 To check the accuracy of extracted $\tau_R$, we progressively 
increase  the number of targeted eigenstates from $100$ to $5500$, which indicates the 
 convergence of $\tau_R$ to its analytical estimation ${\text{\small{$\tau_R t_m\propto(t_e/t_m)^{N_m-1}$}}}$. 
 We also considered different targeted energy densities, centered at $j_e/2$ and $j_e/4$, 
 which does not show any quantitative difference. The reported results are
 averaged over $600$ realizations of initial states and extracted data from targeted 
 energy densities centered at $j_e/2$ and $j_e/4$.
 
%
%
\noindent  {\bf Projected Hamiltonian.}---As mentioned in the main text,
dynamics of the fluxes in our model is governed by the two-body
ferromagnetic Ising interaction, $X_{i} X_{j}$ between the nearest 
neighbor spins, that sit on the legs of the ladder. Here, we would like to
restrict our study to the effect of flux dynamics 
on localization or thermalization of finite energy 
states, therefore we discard the creation/annihilation of fluxes
which might occurs as a result of $X_{i} X_{j}$ terms. 
Typically, this recombination processes are less likely 
for the heavy particles in comparison with the light ones, 
equivalent to the condition $j_{m}\gg t_{e} > t_{m}$.
 \begin{figure}[ht]
   \centering
   \includegraphics[width=3.2in]{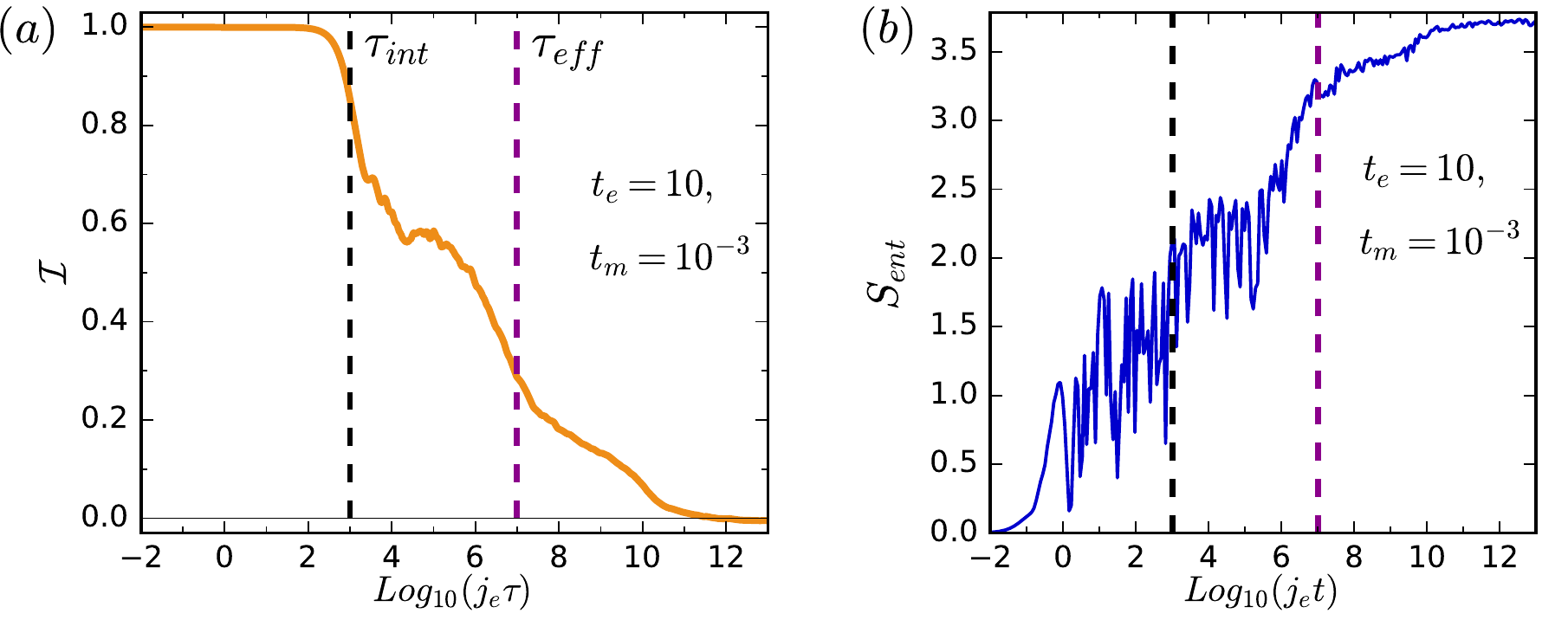}
   \caption{ (a) Relaxation of flux inhomogeneity and (b) Growth of entanglement 
   entropy for subsystem cut at $8$-th spin, corresponding 
   to strong effective disorder governed by $\tilde{H}_{KL}$ for 
   $(N,\,N_m)=(15,\,3)$, averaged over 200 random product states.}
   \label{fig:inhomo_Sent_PNm}
 \end{figure}
Here, we impose this restriction by projecting the leg-Ising interaction onto 
the subspace with a fixed total number of flux by $\hat{P}_{N_{m}}=\sum^{'}_{\{n_{i}^{m}\}} \hat{P}_{\{n_{i}^{m}\}}$, 
where $\hat{P}_{\{n_{i}^{m}\}}$ is defined in Eq.~(\ref{eq:close_form_ev}), 
and the summation is restricted to those configurations in which $N_m=\sum_{i}n_{i}^{m}$ is fixed.
In this respect, the projected Hamiltonian is given by 
$\tilde{H}_{KL}=\hat{P}_{N_{m}} H^{KL}\hat{P}_{N_{m}}$.
The matrix elements of $\tilde{H}_{KL}$ can only 
connect different configurations of the flux anyons with fixed $N_m$. Since
the total number of fluxes, in addition to the flux parity, will be a constant 
of motion of $\tilde{H}_{KL}$, using this projected Hamiltonian is more efficient 
for computational purposes. 
In this situation, the effective mass of heavy particles living in finite 
energy levels is inversely controlled by $t_{m}$ in each fixed
flux sector, $j_{m}$ is an irrelevant parameter.
To check the validity of this approach, in Fig.~\ref{fig:inhomo_Sent_PNm} 
we compute entanglement dynamics and relaxation of the initial inhomogeneity
governed by $\tilde{H}_{KL}$, which shows the same qualitative behavior in 
comparison to those governed by $H^{KL}$ presented in the main text.
 \begin{figure}[t!]
   \centering
   \includegraphics[width=6.2in]{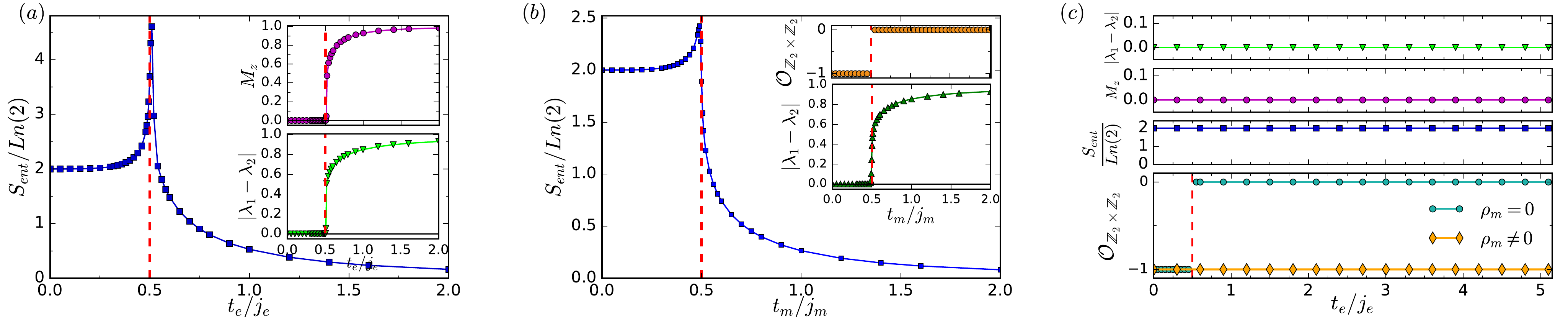}
   \caption{(a) von-Neumann entropy $S_{ent}$, calculated by an implementation 
   of iTEBD method with $\chi=64$ for $H^{KL}(t_m=0)$ in $\rho_{m}=0$ sector.
   The upper and bottom insets show magnetization, $M_{z}$, and the gap between 
two largest Schmidt coefficients $\vert \lambda_{1}-\lambda _{2}\vert$, 
respectively. All data shows diverging behavior at $t_{e}^{c}=j_{e}/2$. 
   (b) von-Neumann entropy computed for $H^{KL}(t_e=0)$ in $\rho_{e}=0$ sector. 
   The upper and bottom insets show $\mathcal{O}_{\mathbb{Z}_{2} \times \mathbb{Z}_{2}}$ defined
   in Eq.~(\ref{eq:PFOP}) and gap between two largest Schmidt coefficients, 
respectively. 
   (c) The three top panels display the effect of nonzero density of fluxes
   on the same quantities plotted in (a). The lowest panel specifies the phase 
factor order parameter
   ,$\mathcal{O}_{\mathbb{Z}_{2} \times \mathbb{Z}_{2}}$, which reveals 
   robustness of the SPT order for $\rho_m \neq 0$.}
   \label{fig:free_flux}
 \end{figure}  
%
%

\noindent {\bf Robustness of SPT phase in finite flux density.}---To reveal 
the effect of nonzero flux density on robustness of the SPT order, 
we consider $H^{KL}_{\rm dual}$ with $t_m=0$. First, in $\rho_m=0$ the
dual Hamiltonian reduces to two decoupled \textit{clean} TFI chains. 
Accordingly, the system has a well-known phase transition
at $t^c_e = j_e/2$. The paramagnetic phase for $t_e<j_e/2$ in the 
$\mu_{r/b}$-picture is identical to the $\mathbb{Z}_2 \times \mathbb{Z}_2$ 
SPT phase in the language of the original model~\cite{Langari:2015}, 
while $t_e > j_e/2$ is smoothly connected to the $j_e=0$ 
fixed point and represents the trivial polarized phase.
However, for $\rho_m>0$ the TFI chains are no longer clean and effectively 
experience a random dilution disorder. According to the the exact 
solution presented in Ref.~\onlinecite{Stinchcombe:1981}, for arbitrary 
(increasing) values of $t_e$, even an infinitesimal $\rho _{m}$ spoils any long range magnetic order. 
For this reason, in the $\mu_{r/b}$-picture system is always in the
paramagnetic phase, and thus the existence of m-errors favors the topological order.

\emph{Numerical approach.}---We employed infinite time-evolving 
block decimation (iTEBD) algorithm~\cite{Orus:2008}, which is based on 
infinite matrix-product state (iMPS) representation to numerically 
check the validity of above analytical results. 
In the original model, for the case of $\rho_{m}=0$, we compute 
the half-cut von-Neumann entanglement entropy,
$S_{ent}= - tr(\rho_{L/2} \ln \rho_{L/2})$, 
where $\rho$ is the ground state density matrix. 
The main plot of Fig.~\ref{fig:free_flux}-(a) shows a diverging behavior of
$S_{ent}$ at $t_{e}^{c}=j_e/2$.
Moreover, the magnetization, $M_z=\sum_i Z_i/N$ (presented in the upper
inset of Fig.~\ref{fig:free_flux}-(a)), shows a transition from a non-magnetic
phase for $t_{e}<j_e/2$ to a spontaneously symmetry breaking phase for 
$t_{e}>j_e/2$.
The lower inset of Fig.~\ref{fig:free_flux}-(a) unveils the difference between 
the two largest magnitudes of Schmidt coefficients, $\vert 
\lambda_{1}-\lambda_{2}\vert$, 
where the degeneracy of $\lambda_1$ and $\lambda_2$ for $t_{e}<j_e/2$ is the 
characteristic feature of SPT phase~\cite{Pollmann:2010}. 
A similar behavior holds in the case $H^{KL}(t_e=0)$ in the
$\rho_{e}=0$ sector, where the phase transition occurs at 
$t_{m}^{c}=j_m/2$ (see Fig~\ref{fig:free_flux}-(b)).
Additionally, we compute the phase factor order parameter~\cite{Pollmann:2012}:
\begin{equation}
\label{eq:PFOP}
\mathcal{O}_{\mathbb{Z}_{2} \times \mathbb{Z}_{2}} = \left\{ \begin{array}{ll}
        0 & \mbox{if $|\eta_{z}|<1$ or $|\eta_{x}|<1$ }\\
        \dfrac{1}{\chi} tr(U_{z} U_{x} U_{z}^{\dagger} U_{x}^{\dagger}) & 
\mbox{if $|\eta_{z}|<1=|\eta_{x}|=1$,}\end{array} \right. 
\end{equation}
where $\eta_{z}$ and $\eta_{x}$ are the largest eigenvalues of the
\textit{generalized transfer matrix}~\cite{Pollmann:2012} constructed by
symmetry operators $\mathcal{P}_m$ and $\mathcal{P}_e$, respectively, 
and $U_{z}^{\dagger}$ and $U_{x}^{\dagger}$ are eigenvectors corresponding to 
them (here $\chi$ is the bond dimension of iMPS).
The symmetry protected nontrivial
phase, symmetry-breaking phase and symmetry protected trivial phase are 
characterized by 
$\mathcal{O}_{\mathbb{Z}_{2} \times \mathbb{Z}_{2}}=\{-1,0,1\}$, respectively. 
The behavior of the mentioned quantities implies that for $t_{e}<j_e/2$ 
($t_{m}<j_m/2$) the perturbed system at
zero flux sector (zero charge sector) belongs to the $\mathbb{Z}_2 \times 
\mathbb{Z}_2$ quantum spin liquid
phase.

Now, we consider the parent Hamiltonian $H_{0}^{'KL}$, 
which is different from $H_{0}^{KL}$ in that, we set
$j_{m}\rightarrow j^{'}_{m}=-j_{m}$. The ground state of $H_{0}^{'KL}$
is in the full flux sector equivalent to the highest energy 
sector of $H_{0}^{KL}$. It can be shown that after adding the 
plaquette-Ising interaction (as charge kinetic term)
to the $H_{0}^{'KL}$, the ground state of perturbed parent Hamiltonian, 
$H^{'}_{KL}$, remains in $\rho _{m}\neq0$ sectors, provided that the rough estimate 
$t_{e}\lesssim j_m/2$ holds. It is worth mentioning that iTEBD algorithm does not
guarantee the final wavefunction of $H^{'}_{KL}$ to remain in a sector 
with finite flux density for arbitrary enhancing value of $t_{e}$.
Hence, mentioned constraint ensures that the zero-flux sector is not reachable as $t_{e}$ is increased.

The results of our numerical simulation for $H^{'}_{KL}$ with 
$j^{'}_{m}=-10j_{e}$ are presented in 
Fig.~\ref{fig:free_flux}-(c), which shows no evidence for a quantum phase 
transition at $\rho _{m}\neq0$. The von-Neumann entropy is $2\ln2$, where magnetization is zero, the Schmidt 
coefficients are degenerate and $\mathcal{O}_{\mathbb{Z}_{2} \times \mathbb{Z}_{2}}=-1$. 
All these results indicate the robustness of SPT order as a consequence of 
self-generated disorder. This effect is 
also comparable with Ref.~\onlinecite{Tsomokos:2011}, in which the \textit{external} 
random field with dilution distribution stabilizes intrinsic topological order
against \textit{arbitrarily} strong magnetic fields.
\begin{figure}[t!]
\centering
\centerline{\includegraphics[width=5.75in]{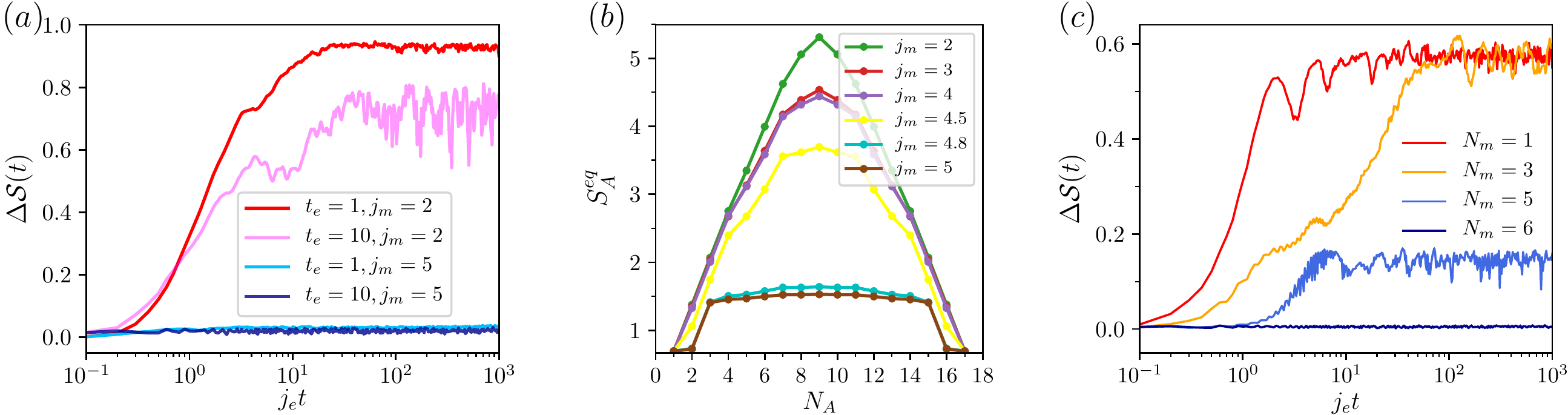}}
\caption{ (a) Spreading of the information for pre-quench state initialized in 
$\rho_m=1$, $t_m=0.6$ and $N=18$. 
(b) Scaling of $S_{ent}$ with different subsystem sizes for $t_e=1$, $t_m=0.6$ and $\rho_m=1$.
(c) The effect of initial flux density on spreading of information for 
$t_e=10$, $t_m=0.3$ and $j_m=5$.
}
\label{fig:advantages}
\end{figure}
%
%

\noindent  {\bf The role of interaction and effective temperature in the quenching procedure.}---Here, we 
investigate the dependence of the non-ergodic quench behavior on the 
interaction and effective temperature.
Apart from the parameters $j_m$ and $t_m$, the flux density in pre-quench state and $t_e$ can be seen 
as additional controlling parameters for initial energy density (trough the relation $\varepsilon_{\rho_{m}} = -j_m (1-\rho_m)-2j_e$) and the strength of the effective disorder,
respectively. By reducing the strength of interaction from $t_e=10$ (see Fig.~\ref{fig:KrylovJZ10V2} in the main text) to $t_e=1$, the information spreading escalates as shown in Fig.~\ref{fig:advantages}-(a).
Subsequently, as shown in Fig.~\ref{fig:advantages}-(b), the scaling of entanglement changes behavior from area law to volume law, approximately around $j_m = 4.5$, which become greater than $j_m \sim 3$ for $t_e = 10$. Moreover, lowering the number of initial fluxes in 
the pre-quench state results in an upsurge in the heating 
process, in the way that the Ising perturbations adversely redesign the initial 
short-range state to a mixed one (see Fig.~\ref{fig:advantages}-(c)).
As a result, increasing the effective temperature and strength of interaction 
ameliorates the resilience of topological order as well as robustness of the initial encoded information.

\noindent  {\bf Generalization to other SPT phases.}---We would like to present 
a more general picture on the implications of topological order and braiding 
statistics for the self-localization of anyon excitations in highly
excited states.  In quantum double models such as Levin-Wen model or Kitaev's 
toric code, an anyon of type $a$ can be transported from site $i$ to site $j$ along 
the directed path $\gamma$ connecting the two sites through applying open 
string operators (Wilson line operators) of type $a$,  $W^a_{\gamma}$. In order to 
mobilize anyons in the system and let them acquire kinetic energy, we can add 
 $\sum_{a}\sum_{\gamma}g^{a}_{\gamma}W^{a}_{\gamma} $ to the ideal
 (exactly solvable) Hamiltonian, where $g^{a}_{\gamma}$ is the amplitude of 
 path $\gamma$. Now, imagine two distinct paths $\gamma_1$ and $\gamma_2$.
 The product of the two open string operators along $\gamma_1$ and 
$\overline{\gamma_2}$ ($\gamma_2$ with opposite direction), i.e.
$W_{c} \equiv W^{a}_{\gamma_1} W^{a}_{\overline{\gamma_2}}=W^{a}_{\gamma_1} \left(W^{a}_{\gamma_2}\right)^{-1}$, 
 forms a closed string (Wilson loop operator). The resulting loop can take 
 various quantized  values depending on the total 
 anyon charge inside path $\gamma_3 \equiv \gamma_1 \overline{\gamma_2}$. 
 Now suppose the resulting path, i.e. $\gamma_3$, encloses total anyon charge equal to $b$. 
 The Wilson operator $W_c$ measures the braid statistics between anyons 
 $a$ and $b$ and is independent of path $\gamma_3$ 
 (as far as it encompasses the anyon charge $b$). Therefore,
 $W^{a}_{\gamma_2} = W_{c}  W^{a}_{\gamma_1}$. 
 As a result, the total contribution of the two paths $\gamma_1$ and
 $\gamma_2$ is $\left( g^a_{\gamma_1} + g^a_{\gamma_2}W_c\right)W^{a}_{\gamma_1}$. 
 Now, let us assume the two paths $\gamma_1$ and $\gamma_2$ are related via some symmetry 
operations, for instance the mirror symmetry with respect to the $x$ axis. In 
that case, we must choose $\left| g^{a}_{\gamma_1}\right|= \left|g^a_{\gamma_2}\right|$ if we 
want to preserve that symmetry. Since, $W_{c}$ depends on the total anyon charge 
trapped inside loop $\gamma_3$ and it may take a different value upon the 
translation of $\gamma_3$, the perturbations can be viewed as disordered anyon hopping terms 
where the disorder is due to the nontrivial anyon braid statistics as discussed 
previously.

\begin{thebibliography}{70}%
\makeatletter
\providecommand \@ifxundefined [1]{%
 \@ifx{#1\undefined}
}%
\providecommand \@ifnum [1]{%
 \ifnum #1\expandafter \@firstoftwo
 \else \expandafter \@secondoftwo
 \fi
}%
\providecommand \@ifx [1]{%
 \ifx #1\expandafter \@firstoftwo
 \else \expandafter \@secondoftwo
 \fi
}%
\providecommand \natexlab [1]{#1}%
\providecommand \enquote  [1]{``#1''}%
\providecommand \bibnamefont  [1]{#1}%
\providecommand \bibfnamefont [1]{#1}%
\providecommand \citenamefont [1]{#1}%
\providecommand \href@noop [0]{\@secondoftwo}%
\providecommand \href [0]{\begingroup \@sanitize@url \@href}%
\providecommand \@href[1]{\@@startlink{#1}\@@href}%
\providecommand \@@href[1]{\endgroup#1\@@endlink}%
\providecommand \@sanitize@url [0]{\catcode `\\12\catcode `\$12\catcode
  `\&12\catcode `\#12\catcode `\^12\catcode `\_12\catcode `\%12\relax}%
\providecommand \@@startlink[1]{}%
\providecommand \@@endlink[0]{}%
\providecommand \url  [0]{\begingroup\@sanitize@url \@url }%
\providecommand \@url [1]{\endgroup\@href {#1}{\urlprefix }}%
\providecommand \urlprefix  [0]{URL }%
\providecommand \Eprint [0]{\href }%
\providecommand \doibase [0]{http://dx.doi.org/}%
\providecommand \selectlanguage [0]{\@gobble}%
\providecommand \bibinfo  [0]{\@secondoftwo}%
\providecommand \bibfield  [0]{\@secondoftwo}%
\providecommand \translation [1]{[#1]}%
\providecommand \BibitemOpen [0]{}%
\providecommand \bibitemStop [0]{}%
\providecommand \bibitemNoStop [0]{.\EOS\space}%
\providecommand \EOS [0]{\spacefactor3000\relax}%
\providecommand \BibitemShut  [1]{\csname bibitem#1\endcsname}%
\let\auto@bib@innerbib\@empty
\bibitem [{\citenamefont {Gornyi}\ \emph {et~al.}(2005)\citenamefont {Gornyi},
  \citenamefont {Mirlin},\ and\ \citenamefont {Polyakov}}]{Gornyi:2005}%
  \BibitemOpen
  \bibfield  {author} {\bibinfo {author} {\bibfnamefont {I.~V.}\ \bibnamefont
  {Gornyi}}, \bibinfo {author} {\bibfnamefont {A.~D.}\ \bibnamefont {Mirlin}},
  \ and\ \bibinfo {author} {\bibfnamefont {D.~G.}\ \bibnamefont {Polyakov}},\
  }\href {\doibase 10.1103/PhysRevLett.95.206603} {\bibfield  {journal}
  {\bibinfo  {journal} {Phys. Rev. Lett.}\ }\textbf {\bibinfo {volume} {95}},\
  \bibinfo {pages} {206603} (\bibinfo {year} {2005})}\BibitemShut {NoStop}%
\bibitem [{\citenamefont {Basko}\ \emph {et~al.}(2006)\citenamefont {Basko},
  \citenamefont {Aleiner},\ and\ \citenamefont {Altshuler}}]{Basko:2006}%
  \BibitemOpen
  \bibfield  {author} {\bibinfo {author} {\bibfnamefont {D.}~\bibnamefont
  {Basko}}, \bibinfo {author} {\bibfnamefont {I.}~\bibnamefont {Aleiner}}, \
  and\ \bibinfo {author} {\bibfnamefont {B.}~\bibnamefont {Altshuler}},\ }\href
  {http://www.sciencedirect.com/science/article/pii/S0003491605002630}
  {\bibfield  {journal} {\bibinfo  {journal} {Annals of Physics}\ }\textbf
  {\bibinfo {volume} {321}},\ \bibinfo {pages} {1126} (\bibinfo {year}
  {2006})}\BibitemShut {NoStop}%
\bibitem [{\citenamefont {Oganesyan}\ and\ \citenamefont
  {Huse}(2007)}]{Oganesyan:2007}%
  \BibitemOpen
  \bibfield  {author} {\bibinfo {author} {\bibfnamefont {V.}~\bibnamefont
  {Oganesyan}}\ and\ \bibinfo {author} {\bibfnamefont {D.~A.}\ \bibnamefont
  {Huse}},\ }\href {\doibase 10.1103/PhysRevB.75.155111} {\bibfield  {journal}
  {\bibinfo  {journal} {Phys. Rev. B}\ }\textbf {\bibinfo {volume} {75}},\
  \bibinfo {pages} {155111} (\bibinfo {year} {2007})}\BibitemShut {NoStop}%
\bibitem [{\citenamefont {Pal}\ and\ \citenamefont {Huse}(2010)}]{Pal:2010}%
  \BibitemOpen
  \bibfield  {author} {\bibinfo {author} {\bibfnamefont {A.}~\bibnamefont
  {Pal}}\ and\ \bibinfo {author} {\bibfnamefont {D.~A.}\ \bibnamefont {Huse}},\
  }\href {\doibase 10.1103/PhysRevB.82.174411} {\bibfield  {journal} {\bibinfo
  {journal} {Phys. Rev. B}\ }\textbf {\bibinfo {volume} {82}},\ \bibinfo
  {pages} {174411} (\bibinfo {year} {2010})}\BibitemShut {NoStop}%
\bibitem [{\citenamefont {Bauer}\ and\ \citenamefont
  {Nayak}(2013)}]{Bauer:2013}%
  \BibitemOpen
  \bibfield  {author} {\bibinfo {author} {\bibfnamefont {B.}~\bibnamefont
  {Bauer}}\ and\ \bibinfo {author} {\bibfnamefont {C.}~\bibnamefont {Nayak}},\
  }\href {http://stacks.iop.org/1742-5468/2013/i=09/a=P09005} {\bibfield
  {journal} {\bibinfo  {journal} {Journal of Statistical Mechanics: Theory and
  Experiment}\ }\textbf {\bibinfo {volume} {2013}},\ \bibinfo {pages} {P09005}
  (\bibinfo {year} {2013})}\BibitemShut {NoStop}%
\bibitem [{\citenamefont {Imbrie}(2016)}]{Imbrie:2016}%
  \BibitemOpen
  \bibfield  {author} {\bibinfo {author} {\bibfnamefont {J.~Z.}\ \bibnamefont
  {Imbrie}},\ }\href {\doibase 10.1007/s10955-016-1508-x} {\bibfield  {journal}
  {\bibinfo  {journal} {Journal of Statistical Physics}\ }\textbf {\bibinfo
  {volume} {163}},\ \bibinfo {pages} {998} (\bibinfo {year}
  {2016})}\BibitemShut {NoStop}%
\bibitem [{\citenamefont {Anderson}(1958)}]{Anderson:1958}%
  \BibitemOpen
  \bibfield  {author} {\bibinfo {author} {\bibfnamefont {P.~W.}\ \bibnamefont
  {Anderson}},\ }\href {\doibase 10.1103/PhysRev.109.1492} {\bibfield
  {journal} {\bibinfo  {journal} {Phys. Rev.}\ }\textbf {\bibinfo {volume}
  {109}},\ \bibinfo {pages} {1492} (\bibinfo {year} {1958})}\BibitemShut
  {NoStop}%
\bibitem [{\citenamefont {Kj\"all}\ \emph {et~al.}(2014)\citenamefont
  {Kj\"all}, \citenamefont {Bardarson},\ and\ \citenamefont
  {Pollmann}}]{Kjall:2014}%
  \BibitemOpen
  \bibfield  {author} {\bibinfo {author} {\bibfnamefont {J.~A.}\ \bibnamefont
  {Kj\"all}}, \bibinfo {author} {\bibfnamefont {J.~H.}\ \bibnamefont
  {Bardarson}}, \ and\ \bibinfo {author} {\bibfnamefont {F.}~\bibnamefont
  {Pollmann}},\ }\href {\doibase 10.1103/PhysRevLett.113.107204} {\bibfield
  {journal} {\bibinfo  {journal} {Phys. Rev. Lett.}\ }\textbf {\bibinfo
  {volume} {113}},\ \bibinfo {pages} {107204} (\bibinfo {year}
  {2014})}\BibitemShut {NoStop}%
\bibitem [{\citenamefont {Serbyn}\ \emph
  {et~al.}(2014{\natexlab{a}})\citenamefont {Serbyn}, \citenamefont
  {Papi\ifmmode~\acute{c}\else \'{c}\fi{}},\ and\ \citenamefont
  {Abanin}}]{Serbyn:2014_1}%
  \BibitemOpen
  \bibfield  {author} {\bibinfo {author} {\bibfnamefont {M.}~\bibnamefont
  {Serbyn}}, \bibinfo {author} {\bibfnamefont {Z.}~\bibnamefont
  {Papi\ifmmode~\acute{c}\else \'{c}\fi{}}}, \ and\ \bibinfo {author}
  {\bibfnamefont {D.~A.}\ \bibnamefont {Abanin}},\ }\href {\doibase
  10.1103/PhysRevB.90.174302} {\bibfield  {journal} {\bibinfo  {journal} {Phys.
  Rev. B}\ }\textbf {\bibinfo {volume} {90}},\ \bibinfo {pages} {174302}
  (\bibinfo {year} {2014}{\natexlab{a}})}\BibitemShut {NoStop}%
\bibitem [{\citenamefont {Serbyn}\ \emph
  {et~al.}(2014{\natexlab{b}})\citenamefont {Serbyn}, \citenamefont {Knap},
  \citenamefont {Gopalakrishnan}, \citenamefont {Papi\ifmmode~\acute{c}\else
  \'{c}\fi{}}, \citenamefont {Yao}, \citenamefont {Laumann}, \citenamefont
  {Abanin}, \citenamefont {Lukin},\ and\ \citenamefont
  {Demler}}]{Serbyn:2014_2}%
  \BibitemOpen
  \bibfield  {author} {\bibinfo {author} {\bibfnamefont {M.}~\bibnamefont
  {Serbyn}}, \bibinfo {author} {\bibfnamefont {M.}~\bibnamefont {Knap}},
  \bibinfo {author} {\bibfnamefont {S.}~\bibnamefont {Gopalakrishnan}},
  \bibinfo {author} {\bibfnamefont {Z.}~\bibnamefont
  {Papi\ifmmode~\acute{c}\else \'{c}\fi{}}}, \bibinfo {author} {\bibfnamefont
  {N.~Y.}\ \bibnamefont {Yao}}, \bibinfo {author} {\bibfnamefont {C.~R.}\
  \bibnamefont {Laumann}}, \bibinfo {author} {\bibfnamefont {D.~A.}\
  \bibnamefont {Abanin}}, \bibinfo {author} {\bibfnamefont {M.~D.}\
  \bibnamefont {Lukin}}, \ and\ \bibinfo {author} {\bibfnamefont {E.~A.}\
  \bibnamefont {Demler}},\ }\href {\doibase 10.1103/PhysRevLett.113.147204}
  {\bibfield  {journal} {\bibinfo  {journal} {Phys. Rev. Lett.}\ }\textbf
  {\bibinfo {volume} {113}},\ \bibinfo {pages} {147204} (\bibinfo {year}
  {2014}{\natexlab{b}})}\BibitemShut {NoStop}%
\bibitem [{\citenamefont {\ifmmode \check{Z}\else
  \v{Z}\fi{}nidari\ifmmode~\check{c}\else \v{c}\fi{}}\ \emph
  {et~al.}(2008)\citenamefont {\ifmmode \check{Z}\else
  \v{Z}\fi{}nidari\ifmmode~\check{c}\else \v{c}\fi{}}, \citenamefont {Prosen},\
  and\ \citenamefont {Prelov\ifmmode~\check{s}\else
  \v{s}\fi{}ek}}]{Znidaric:2008}%
  \BibitemOpen
  \bibfield  {author} {\bibinfo {author} {\bibfnamefont {M.}~\bibnamefont
  {\ifmmode \check{Z}\else \v{Z}\fi{}nidari\ifmmode~\check{c}\else
  \v{c}\fi{}}}, \bibinfo {author} {\bibfnamefont {T.~c.~v.}\ \bibnamefont
  {Prosen}}, \ and\ \bibinfo {author} {\bibfnamefont {P.}~\bibnamefont
  {Prelov\ifmmode~\check{s}\else \v{s}\fi{}ek}},\ }\href {\doibase
  10.1103/PhysRevB.77.064426} {\bibfield  {journal} {\bibinfo  {journal} {Phys.
  Rev. B}\ }\textbf {\bibinfo {volume} {77}},\ \bibinfo {pages} {064426}
  (\bibinfo {year} {2008})}\BibitemShut {NoStop}%
\bibitem [{\citenamefont {Bardarson}\ \emph {et~al.}(2012)\citenamefont
  {Bardarson}, \citenamefont {Pollmann},\ and\ \citenamefont
  {Moore}}]{Bardarson:2012}%
  \BibitemOpen
  \bibfield  {author} {\bibinfo {author} {\bibfnamefont {J.~H.}\ \bibnamefont
  {Bardarson}}, \bibinfo {author} {\bibfnamefont {F.}~\bibnamefont {Pollmann}},
  \ and\ \bibinfo {author} {\bibfnamefont {J.~E.}\ \bibnamefont {Moore}},\
  }\href {\doibase 10.1103/PhysRevLett.109.017202} {\bibfield  {journal}
  {\bibinfo  {journal} {Phys. Rev. Lett.}\ }\textbf {\bibinfo {volume} {109}},\
  \bibinfo {pages} {017202} (\bibinfo {year} {2012})}\BibitemShut {NoStop}%
\bibitem [{\citenamefont {Vosk}\ and\ \citenamefont
  {Altman}(2013)}]{Vosk:2013}%
  \BibitemOpen
  \bibfield  {author} {\bibinfo {author} {\bibfnamefont {R.}~\bibnamefont
  {Vosk}}\ and\ \bibinfo {author} {\bibfnamefont {E.}~\bibnamefont {Altman}},\
  }\href {\doibase 10.1103/PhysRevLett.110.067204} {\bibfield  {journal}
  {\bibinfo  {journal} {Phys. Rev. Lett.}\ }\textbf {\bibinfo {volume} {110}},\
  \bibinfo {pages} {067204} (\bibinfo {year} {2013})}\BibitemShut {NoStop}%
\bibitem [{\citenamefont {Serbyn}\ \emph
  {et~al.}(2013{\natexlab{a}})\citenamefont {Serbyn}, \citenamefont
  {Papi\ifmmode~\acute{c}\else \'{c}\fi{}},\ and\ \citenamefont
  {Abanin}}]{Serbyn:2013_1}%
  \BibitemOpen
  \bibfield  {author} {\bibinfo {author} {\bibfnamefont {M.}~\bibnamefont
  {Serbyn}}, \bibinfo {author} {\bibfnamefont {Z.}~\bibnamefont
  {Papi\ifmmode~\acute{c}\else \'{c}\fi{}}}, \ and\ \bibinfo {author}
  {\bibfnamefont {D.~A.}\ \bibnamefont {Abanin}},\ }\href {\doibase
  10.1103/PhysRevLett.110.260601} {\bibfield  {journal} {\bibinfo  {journal}
  {Phys. Rev. Lett.}\ }\textbf {\bibinfo {volume} {110}},\ \bibinfo {pages}
  {260601} (\bibinfo {year} {2013}{\natexlab{a}})}\BibitemShut {NoStop}%
\bibitem [{\citenamefont {Deutsch}(1991)}]{Deutsch:1991}%
  \BibitemOpen
  \bibfield  {author} {\bibinfo {author} {\bibfnamefont {J.~M.}\ \bibnamefont
  {Deutsch}},\ }\href {\doibase 10.1103/PhysRevA.43.2046} {\bibfield  {journal}
  {\bibinfo  {journal} {Phys. Rev. A}\ }\textbf {\bibinfo {volume} {43}},\
  \bibinfo {pages} {2046} (\bibinfo {year} {1991})}\BibitemShut {NoStop}%
\bibitem [{\citenamefont {Srednicki}(1994)}]{Srednicki:1994}%
  \BibitemOpen
  \bibfield  {author} {\bibinfo {author} {\bibfnamefont {M.}~\bibnamefont
  {Srednicki}},\ }\href {\doibase 10.1103/PhysRevE.50.888} {\bibfield
  {journal} {\bibinfo  {journal} {Phys. Rev. E}\ }\textbf {\bibinfo {volume}
  {50}},\ \bibinfo {pages} {888} (\bibinfo {year} {1994})}\BibitemShut
  {NoStop}%
\bibitem [{\citenamefont {Rigol}\ \emph {et~al.}(2008)\citenamefont {Rigol},
  \citenamefont {Dunjko},\ and\ \citenamefont {Olshanii}}]{Rigol:2008}%
  \BibitemOpen
  \bibfield  {author} {\bibinfo {author} {\bibfnamefont {M.}~\bibnamefont
  {Rigol}}, \bibinfo {author} {\bibfnamefont {V.}~\bibnamefont {Dunjko}}, \
  and\ \bibinfo {author} {\bibfnamefont {M.}~\bibnamefont {Olshanii}},\ }\href
  {\doibase 10.1038/nature06838} {\bibfield  {journal} {\bibinfo  {journal}
  {Nature}\ }\textbf {\bibinfo {volume} {452}},\ \bibinfo {pages} {854}
  (\bibinfo {year} {2008})}\BibitemShut {NoStop}%
\bibitem [{\citenamefont {Huse}\ \emph {et~al.}(2013)\citenamefont {Huse},
  \citenamefont {Nandkishore}, \citenamefont {Oganesyan}, \citenamefont {Pal},\
  and\ \citenamefont {Sondhi}}]{Huse:2013}%
  \BibitemOpen
  \bibfield  {author} {\bibinfo {author} {\bibfnamefont {D.~A.}\ \bibnamefont
  {Huse}}, \bibinfo {author} {\bibfnamefont {R.}~\bibnamefont {Nandkishore}},
  \bibinfo {author} {\bibfnamefont {V.}~\bibnamefont {Oganesyan}}, \bibinfo
  {author} {\bibfnamefont {A.}~\bibnamefont {Pal}}, \ and\ \bibinfo {author}
  {\bibfnamefont {S.~L.}\ \bibnamefont {Sondhi}},\ }\href {\doibase
  10.1103/PhysRevB.88.014206} {\bibfield  {journal} {\bibinfo  {journal} {Phys.
  Rev. B}\ }\textbf {\bibinfo {volume} {88}},\ \bibinfo {pages} {014206}
  (\bibinfo {year} {2013})}\BibitemShut {NoStop}%
\bibitem [{\citenamefont {Chandran}\ \emph {et~al.}(2014)\citenamefont
  {Chandran}, \citenamefont {Khemani}, \citenamefont {Laumann},\ and\
  \citenamefont {Sondhi}}]{Chandran:2014}%
  \BibitemOpen
  \bibfield  {author} {\bibinfo {author} {\bibfnamefont {A.}~\bibnamefont
  {Chandran}}, \bibinfo {author} {\bibfnamefont {V.}~\bibnamefont {Khemani}},
  \bibinfo {author} {\bibfnamefont {C.~R.}\ \bibnamefont {Laumann}}, \ and\
  \bibinfo {author} {\bibfnamefont {S.~L.}\ \bibnamefont {Sondhi}},\ }\href
  {\doibase 10.1103/PhysRevB.89.144201} {\bibfield  {journal} {\bibinfo
  {journal} {Phys. Rev. B}\ }\textbf {\bibinfo {volume} {89}},\ \bibinfo
  {pages} {144201} (\bibinfo {year} {2014})}\BibitemShut {NoStop}%
\bibitem [{\citenamefont {Bahri}\ \emph {et~al.}(2015)\citenamefont {Bahri},
  \citenamefont {Vosk}, \citenamefont {Altman},\ and\ \citenamefont
  {Vishwanath}}]{Bahri:2015}%
  \BibitemOpen
  \bibfield  {author} {\bibinfo {author} {\bibfnamefont {Y.}~\bibnamefont
  {Bahri}}, \bibinfo {author} {\bibfnamefont {R.}~\bibnamefont {Vosk}},
  \bibinfo {author} {\bibfnamefont {E.}~\bibnamefont {Altman}}, \ and\ \bibinfo
  {author} {\bibfnamefont {A.}~\bibnamefont {Vishwanath}},\ }\href {\doibase
  10.1038/ncomms8341} {\bibfield  {journal} {\bibinfo  {journal} {Nature
  Communications}\ }\textbf {\bibinfo {volume} {6}},\ \bibinfo {pages} {7341}
  (\bibinfo {year} {2015})}\BibitemShut {NoStop}%
\bibitem [{\citenamefont {{Potter}}\ and\ \citenamefont
  {{Vishwanath}}(2015)}]{Potter:2015}%
  \BibitemOpen
  \bibfield  {author} {\bibinfo {author} {\bibfnamefont {A.~C.}\ \bibnamefont
  {{Potter}}}\ and\ \bibinfo {author} {\bibfnamefont {A.}~\bibnamefont
  {{Vishwanath}}},\ }\href@noop {} {\bibfield  {journal} {\bibinfo  {journal}
  {ArXiv e-prints}\ } (\bibinfo {year} {2015})},\ \Eprint
  {http://arxiv.org/abs/1506.00592} {arXiv:1506.00592 [cond-mat.dis-nn]}
  \BibitemShut {NoStop}%
\bibitem [{\citenamefont {{Yao}}\ \emph {et~al.}(2015)\citenamefont {{Yao}},
  \citenamefont {{Laumann}},\ and\ \citenamefont {{Vishwanath}}}]{Yao:2015}%
  \BibitemOpen
  \bibfield  {author} {\bibinfo {author} {\bibfnamefont {N.~Y.}\ \bibnamefont
  {{Yao}}}, \bibinfo {author} {\bibfnamefont {C.~R.}\ \bibnamefont
  {{Laumann}}}, \ and\ \bibinfo {author} {\bibfnamefont {A.}~\bibnamefont
  {{Vishwanath}}},\ }\href@noop {} {\bibfield  {journal} {\bibinfo  {journal}
  {ArXiv e-prints}\ } (\bibinfo {year} {2015})},\ \Eprint
  {http://arxiv.org/abs/1508.06995} {arXiv:1508.06995 [quant-ph]} \BibitemShut
  {NoStop}%
\bibitem [{\citenamefont {Carleo~G}(2011)}]{Carleo:2011}%
  \BibitemOpen
  \bibfield  {author} {\bibinfo {author} {\bibfnamefont {S.~M. F.~M.}\
  \bibnamefont {Carleo~G}, \bibfnamefont {Becca~F}},\ }\href
  {http://www.ncbi.nlm.nih.gov/pmc/articles/PMC3272662/} {\bibfield  {journal}
  {\bibinfo  {journal} {Scientific Reports}\ }\textbf {\bibinfo {volume} {2}},\
  \bibinfo {pages} {243} (\bibinfo {year} {2011})}\BibitemShut {NoStop}%
\bibitem [{\citenamefont {{De Roeck}}\ and\ \citenamefont
  {Huveneers}(2014)}]{DeRoeck:2014_1}%
  \BibitemOpen
  \bibfield  {author} {\bibinfo {author} {\bibfnamefont {W.}~\bibnamefont {{De
  Roeck}}}\ and\ \bibinfo {author} {\bibfnamefont {F.}~\bibnamefont
  {Huveneers}},\ }\href {\doibase 10.1007/s00220-014-2116-8} {\bibfield
  {journal} {\bibinfo  {journal} {Communications in Mathematical Physics}\
  }\textbf {\bibinfo {volume} {332}},\ \bibinfo {pages} {1017–1082} (\bibinfo
  {year} {2014})}\BibitemShut {NoStop}%
\bibitem [{\citenamefont {De~Roeck}\ and\ \citenamefont
  {Huveneers}(2014)}]{DeRoeck:2014_2}%
  \BibitemOpen
  \bibfield  {author} {\bibinfo {author} {\bibfnamefont {W.}~\bibnamefont
  {De~Roeck}}\ and\ \bibinfo {author} {\bibfnamefont {F.~m.~c.}\ \bibnamefont
  {Huveneers}},\ }\href {\doibase 10.1103/PhysRevB.90.165137} {\bibfield
  {journal} {\bibinfo  {journal} {Phys. Rev. B}\ }\textbf {\bibinfo {volume}
  {90}},\ \bibinfo {pages} {165137} (\bibinfo {year} {2014})}\BibitemShut
  {NoStop}%
\bibitem [{\citenamefont {Hickey}\ \emph {et~al.}(2016)\citenamefont {Hickey},
  \citenamefont {Genway},\ and\ \citenamefont {Garrahan}}]{Hickey:2016}%
  \BibitemOpen
  \bibfield  {author} {\bibinfo {author} {\bibfnamefont {J.~M.}\ \bibnamefont
  {Hickey}}, \bibinfo {author} {\bibfnamefont {S.}~\bibnamefont {Genway}}, \
  and\ \bibinfo {author} {\bibfnamefont {J.~P.}\ \bibnamefont {Garrahan}},\
  }\href {http://stacks.iop.org/1742-5468/2016/i=5/a=054047} {\bibfield
  {journal} {\bibinfo  {journal} {Journal of Statistical Mechanics: Theory and
  Experiment}\ }\textbf {\bibinfo {volume} {2016}},\ \bibinfo {pages} {054047}
  (\bibinfo {year} {2016})}\BibitemShut {NoStop}%
\bibitem [{\citenamefont {Schiulaz}\ and\ \citenamefont
  {M\"uller}(2014)}]{Schiulaz:2014}%
  \BibitemOpen
  \bibfield  {author} {\bibinfo {author} {\bibfnamefont {M.}~\bibnamefont
  {Schiulaz}}\ and\ \bibinfo {author} {\bibfnamefont {M.}~\bibnamefont
  {M\"uller}},\ }\href {\doibase http://dx.doi.org/10.1063/1.4893505}
  {\bibfield  {journal} {\bibinfo  {journal} {AIP Conference Proceedings}\
  }\textbf {\bibinfo {volume} {1610}},\ \bibinfo {pages} {11} (\bibinfo {year}
  {2014})}\BibitemShut {NoStop}%
\bibitem [{\citenamefont {Schiulaz}\ \emph {et~al.}(2015)\citenamefont
  {Schiulaz}, \citenamefont {Silva},\ and\ \citenamefont
  {M\"uller}}]{Schiulaz:2015}%
  \BibitemOpen
  \bibfield  {author} {\bibinfo {author} {\bibfnamefont {M.}~\bibnamefont
  {Schiulaz}}, \bibinfo {author} {\bibfnamefont {A.}~\bibnamefont {Silva}}, \
  and\ \bibinfo {author} {\bibfnamefont {M.}~\bibnamefont {M\"uller}},\ }\href
  {\doibase 10.1103/PhysRevB.91.184202} {\bibfield  {journal} {\bibinfo
  {journal} {Phys. Rev. B}\ }\textbf {\bibinfo {volume} {91}},\ \bibinfo
  {pages} {184202} (\bibinfo {year} {2015})}\BibitemShut {NoStop}%
\bibitem [{\citenamefont {Papi\'{c}}\ \emph {et~al.}(2015)\citenamefont
  {Papi\'{c}}, \citenamefont {Stoudenmire},\ and\ \citenamefont
  {Abanin}}]{Papic:2015}%
  \BibitemOpen
  \bibfield  {author} {\bibinfo {author} {\bibfnamefont {Z.}~\bibnamefont
  {Papi\'{c}}}, \bibinfo {author} {\bibfnamefont {E.~M.}\ \bibnamefont
  {Stoudenmire}}, \ and\ \bibinfo {author} {\bibfnamefont {D.~A.}\ \bibnamefont
  {Abanin}},\ }\href {\doibase http://dx.doi.org/10.1016/j.aop.2015.08.024}
  {\bibfield  {journal} {\bibinfo  {journal} {Annals of Physics}\ }\textbf
  {\bibinfo {volume} {362}},\ \bibinfo {pages} {714} (\bibinfo {year}
  {2015})}\BibitemShut {NoStop}%
\bibitem [{\citenamefont {Yao}\ \emph {et~al.}(2016)\citenamefont {Yao},
  \citenamefont {Laumann}, \citenamefont {Cirac}, \citenamefont {Lukin},\ and\
  \citenamefont {Moore}}]{Yao:2016}%
  \BibitemOpen
  \bibfield  {author} {\bibinfo {author} {\bibfnamefont {N.~Y.}\ \bibnamefont
  {Yao}}, \bibinfo {author} {\bibfnamefont {C.~R.}\ \bibnamefont {Laumann}},
  \bibinfo {author} {\bibfnamefont {J.~I.}\ \bibnamefont {Cirac}}, \bibinfo
  {author} {\bibfnamefont {M.~D.}\ \bibnamefont {Lukin}}, \ and\ \bibinfo
  {author} {\bibfnamefont {J.~E.}\ \bibnamefont {Moore}},\ }\href {\doibase
  10.1103/PhysRevLett.117.240601} {\bibfield  {journal} {\bibinfo  {journal}
  {Phys. Rev. Lett.}\ }\textbf {\bibinfo {volume} {117}},\ \bibinfo {pages}
  {240601} (\bibinfo {year} {2016})}\BibitemShut {NoStop}%
\bibitem [{\citenamefont {Barbiero}\ \emph {et~al.}(2015)\citenamefont
  {Barbiero}, \citenamefont {Menotti}, \citenamefont {Recati},\ and\
  \citenamefont {Santos}}]{Barbiero:2015}%
  \BibitemOpen
  \bibfield  {author} {\bibinfo {author} {\bibfnamefont {L.}~\bibnamefont
  {Barbiero}}, \bibinfo {author} {\bibfnamefont {C.}~\bibnamefont {Menotti}},
  \bibinfo {author} {\bibfnamefont {A.}~\bibnamefont {Recati}}, \ and\ \bibinfo
  {author} {\bibfnamefont {L.}~\bibnamefont {Santos}},\ }\href {\doibase
  10.1103/PhysRevB.92.180406} {\bibfield  {journal} {\bibinfo  {journal} {Phys.
  Rev. B}\ }\textbf {\bibinfo {volume} {92}},\ \bibinfo {pages} {180406}
  (\bibinfo {year} {2015})}\BibitemShut {NoStop}%
\bibitem [{\citenamefont {Smith}\ \emph {et~al.}(2017)\citenamefont {Smith},
  \citenamefont {Knolle}, \citenamefont {Kovrizhin},\ and\ \citenamefont
  {Moessner}}]{Smith:2017}%
  \BibitemOpen
  \bibfield  {author} {\bibinfo {author} {\bibfnamefont {A.}~\bibnamefont
  {Smith}}, \bibinfo {author} {\bibfnamefont {J.}~\bibnamefont {Knolle}},
  \bibinfo {author} {\bibfnamefont {D.~L.}\ \bibnamefont {Kovrizhin}}, \ and\
  \bibinfo {author} {\bibfnamefont {R.}~\bibnamefont {Moessner}},\ }\href
  {\doibase 10.1103/PhysRevLett.118.266601} {\bibfield  {journal} {\bibinfo
  {journal} {Phys. Rev. Lett.}\ }\textbf {\bibinfo {volume} {118}},\ \bibinfo
  {pages} {266601} (\bibinfo {year} {2017})}\BibitemShut {NoStop}%
\bibitem [{\citenamefont {Serbyn}\ \emph
  {et~al.}(2013{\natexlab{b}})\citenamefont {Serbyn}, \citenamefont
  {Papi\ifmmode~\acute{c}\else \'{c}\fi{}},\ and\ \citenamefont
  {Abanin}}]{Serbyn:2013_2}%
  \BibitemOpen
  \bibfield  {author} {\bibinfo {author} {\bibfnamefont {M.}~\bibnamefont
  {Serbyn}}, \bibinfo {author} {\bibfnamefont {Z.}~\bibnamefont
  {Papi\ifmmode~\acute{c}\else \'{c}\fi{}}}, \ and\ \bibinfo {author}
  {\bibfnamefont {D.~A.}\ \bibnamefont {Abanin}},\ }\href {\doibase
  10.1103/PhysRevLett.111.127201} {\bibfield  {journal} {\bibinfo  {journal}
  {Phys. Rev. Lett.}\ }\textbf {\bibinfo {volume} {111}},\ \bibinfo {pages}
  {127201} (\bibinfo {year} {2013}{\natexlab{b}})}\BibitemShut {NoStop}%
\bibitem [{\citenamefont {Huse}\ \emph {et~al.}(2014)\citenamefont {Huse},
  \citenamefont {Nandkishore},\ and\ \citenamefont {Oganesyan}}]{Huse:2014}%
  \BibitemOpen
  \bibfield  {author} {\bibinfo {author} {\bibfnamefont {D.~A.}\ \bibnamefont
  {Huse}}, \bibinfo {author} {\bibfnamefont {R.}~\bibnamefont {Nandkishore}}, \
  and\ \bibinfo {author} {\bibfnamefont {V.}~\bibnamefont {Oganesyan}},\ }\href
  {\doibase 10.1103/PhysRevB.90.174202} {\bibfield  {journal} {\bibinfo
  {journal} {Phys. Rev. B}\ }\textbf {\bibinfo {volume} {90}},\ \bibinfo
  {pages} {174202} (\bibinfo {year} {2014})}\BibitemShut {NoStop}%
\bibitem [{\citenamefont {Chandran}\ \emph {et~al.}(2015)\citenamefont
  {Chandran}, \citenamefont {Kim}, \citenamefont {Vidal},\ and\ \citenamefont
  {Abanin}}]{Chandran:2015}%
  \BibitemOpen
  \bibfield  {author} {\bibinfo {author} {\bibfnamefont {A.}~\bibnamefont
  {Chandran}}, \bibinfo {author} {\bibfnamefont {I.~H.}\ \bibnamefont {Kim}},
  \bibinfo {author} {\bibfnamefont {G.}~\bibnamefont {Vidal}}, \ and\ \bibinfo
  {author} {\bibfnamefont {D.~A.}\ \bibnamefont {Abanin}},\ }\href {\doibase
  10.1103/PhysRevB.91.085425} {\bibfield  {journal} {\bibinfo  {journal} {Phys.
  Rev. B}\ }\textbf {\bibinfo {volume} {91}},\ \bibinfo {pages} {085425}
  (\bibinfo {year} {2015})}\BibitemShut {NoStop}%
\bibitem [{\citenamefont {Ros}\ \emph {et~al.}(2015)\citenamefont {Ros},
  \citenamefont {M\"uller},\ and\ \citenamefont {Scardicchio}}]{Ros:2015}%
  \BibitemOpen
  \bibfield  {author} {\bibinfo {author} {\bibfnamefont {V.}~\bibnamefont
  {Ros}}, \bibinfo {author} {\bibfnamefont {M.}~\bibnamefont {M\"uller}}, \
  and\ \bibinfo {author} {\bibfnamefont {A.}~\bibnamefont {Scardicchio}},\
  }\href {http://www.sciencedirect.com/science/article/pii/S0550321314003836}
  {\bibfield  {journal} {\bibinfo  {journal} {Nuclear Physics B}\ }\textbf
  {\bibinfo {volume} {891}},\ \bibinfo {pages} {420} (\bibinfo {year}
  {2015})}\BibitemShut {NoStop}%
\bibitem [{\citenamefont {Dennis}\ \emph {et~al.}(2002)\citenamefont {Dennis},
  \citenamefont {Kitaev}, \citenamefont {Landahl},\ and\ \citenamefont
  {Preskill}}]{Dennis:2002}%
  \BibitemOpen
  \bibfield  {author} {\bibinfo {author} {\bibfnamefont {E.}~\bibnamefont
  {Dennis}}, \bibinfo {author} {\bibfnamefont {A.}~\bibnamefont {Kitaev}},
  \bibinfo {author} {\bibfnamefont {A.}~\bibnamefont {Landahl}}, \ and\
  \bibinfo {author} {\bibfnamefont {J.}~\bibnamefont {Preskill}},\ }\href
  {\doibase http://dx.doi.org/10.1063/1.1499754} {\bibfield  {journal}
  {\bibinfo  {journal} {Journal of Mathematical Physics}\ }\textbf {\bibinfo
  {volume} {43}},\ \bibinfo {pages} {4452} (\bibinfo {year}
  {2002})}\BibitemShut {NoStop}%
\bibitem [{\citenamefont {Kitaev}(2003)}]{Kitaev:2003}%
  \BibitemOpen
  \bibfield  {author} {\bibinfo {author} {\bibfnamefont {A.}~\bibnamefont
  {Kitaev}},\ }\href {\doibase http://dx.doi.org/10.1016/S0003-4916(02)00018-0}
  {\bibfield  {journal} {\bibinfo  {journal} {Annals of Physics}\ }\textbf
  {\bibinfo {volume} {303}},\ \bibinfo {pages} {2 } (\bibinfo {year}
  {2003})}\BibitemShut {NoStop}%
\bibitem [{\citenamefont {Karimipour}(2009)}]{Karimipour:2009}%
  \BibitemOpen
  \bibfield  {author} {\bibinfo {author} {\bibfnamefont {V.}~\bibnamefont
  {Karimipour}},\ }\href {\doibase 10.1103/PhysRevB.79.214435} {\bibfield
  {journal} {\bibinfo  {journal} {Phys. Rev. B}\ }\textbf {\bibinfo {volume}
  {79}},\ \bibinfo {pages} {214435} (\bibinfo {year} {2009})}\BibitemShut
  {NoStop}%
\bibitem [{\citenamefont {Langari}\ \emph {et~al.}(2015)\citenamefont
  {Langari}, \citenamefont {Mohammad-Aghaei},\ and\ \citenamefont
  {Haghshenas}}]{Langari:2015}%
  \BibitemOpen
  \bibfield  {author} {\bibinfo {author} {\bibfnamefont {A.}~\bibnamefont
  {Langari}}, \bibinfo {author} {\bibfnamefont {A.}~\bibnamefont
  {Mohammad-Aghaei}}, \ and\ \bibinfo {author} {\bibfnamefont {R.}~\bibnamefont
  {Haghshenas}},\ }\href {\doibase 10.1103/PhysRevB.91.024415} {\bibfield
  {journal} {\bibinfo  {journal} {Phys. Rev. B}\ }\textbf {\bibinfo {volume}
  {91}},\ \bibinfo {pages} {024415} (\bibinfo {year} {2015})}\BibitemShut
  {NoStop}%
\bibitem [{\citenamefont {Castelnovo}\ and\ \citenamefont
  {Chamon}(2007)}]{Castelnovo:2007}%
  \BibitemOpen
  \bibfield  {author} {\bibinfo {author} {\bibfnamefont {C.}~\bibnamefont
  {Castelnovo}}\ and\ \bibinfo {author} {\bibfnamefont {C.}~\bibnamefont
  {Chamon}},\ }\href {\doibase 10.1103/PhysRevB.76.184442} {\bibfield
  {journal} {\bibinfo  {journal} {Phys. Rev. B}\ }\textbf {\bibinfo {volume}
  {76}},\ \bibinfo {pages} {184442} (\bibinfo {year} {2007})}\BibitemShut
  {NoStop}%
\bibitem [{\citenamefont {Castelnovo}\ and\ \citenamefont
  {Chamon}(2008)}]{Castelnovo:2008}%
  \BibitemOpen
  \bibfield  {author} {\bibinfo {author} {\bibfnamefont {C.}~\bibnamefont
  {Castelnovo}}\ and\ \bibinfo {author} {\bibfnamefont {C.}~\bibnamefont
  {Chamon}},\ }\href {\doibase 10.1103/PhysRevB.77.054433} {\bibfield
  {journal} {\bibinfo  {journal} {Phys. Rev. B}\ }\textbf {\bibinfo {volume}
  {77}},\ \bibinfo {pages} {054433} (\bibinfo {year} {2008})}\BibitemShut
  {NoStop}%
\bibitem [{\citenamefont {Brown}\ \emph {et~al.}(2016)\citenamefont {Brown},
  \citenamefont {Loss}, \citenamefont {Pachos}, \citenamefont {Self},\ and\
  \citenamefont {Wootton}}]{Brown:2016}%
  \BibitemOpen
  \bibfield  {author} {\bibinfo {author} {\bibfnamefont {B.~J.}\ \bibnamefont
  {Brown}}, \bibinfo {author} {\bibfnamefont {D.}~\bibnamefont {Loss}},
  \bibinfo {author} {\bibfnamefont {J.~K.}\ \bibnamefont {Pachos}}, \bibinfo
  {author} {\bibfnamefont {C.~N.}\ \bibnamefont {Self}}, \ and\ \bibinfo
  {author} {\bibfnamefont {J.~R.}\ \bibnamefont {Wootton}},\ }\href {\doibase
  10.1103/RevModPhys.88.045005} {\bibfield  {journal} {\bibinfo  {journal}
  {Rev. Mod. Phys.}\ }\textbf {\bibinfo {volume} {88}},\ \bibinfo {pages}
  {045005} (\bibinfo {year} {2016})}\BibitemShut {NoStop}%
\bibitem [{\citenamefont {Nussinov}\ and\ \citenamefont
  {Ortiz}(2008)}]{Nussinov:2008}%
  \BibitemOpen
  \bibfield  {author} {\bibinfo {author} {\bibfnamefont {Z.}~\bibnamefont
  {Nussinov}}\ and\ \bibinfo {author} {\bibfnamefont {G.}~\bibnamefont
  {Ortiz}},\ }\href {\doibase 10.1103/PhysRevB.77.064302} {\bibfield  {journal}
  {\bibinfo  {journal} {Phys. Rev. B}\ }\textbf {\bibinfo {volume} {77}},\
  \bibinfo {pages} {064302} (\bibinfo {year} {2008})}\BibitemShut {NoStop}%
\bibitem [{\citenamefont {Hastings}(2011)}]{Hastings:2011}%
  \BibitemOpen
  \bibfield  {author} {\bibinfo {author} {\bibfnamefont {M.~B.}\ \bibnamefont
  {Hastings}},\ }\href {\doibase 10.1103/PhysRevLett.107.210501} {\bibfield
  {journal} {\bibinfo  {journal} {Phys. Rev. Lett.}\ }\textbf {\bibinfo
  {volume} {107}},\ \bibinfo {pages} {210501} (\bibinfo {year}
  {2011})}\BibitemShut {NoStop}%
\bibitem [{\citenamefont {Kay}(2009)}]{Kay:2009}%
  \BibitemOpen
  \bibfield  {author} {\bibinfo {author} {\bibfnamefont {A.}~\bibnamefont
  {Kay}},\ }\href {\doibase 10.1103/PhysRevLett.102.070503} {\bibfield
  {journal} {\bibinfo  {journal} {Phys. Rev. Lett.}\ }\textbf {\bibinfo
  {volume} {102}},\ \bibinfo {pages} {070503} (\bibinfo {year}
  {2009})}\BibitemShut {NoStop}%
\bibitem [{\citenamefont {Zeng}\ \emph {et~al.}(2016)\citenamefont {Zeng},
  \citenamefont {Hamma},\ and\ \citenamefont {Fan}}]{Zeng:2016}%
  \BibitemOpen
  \bibfield  {author} {\bibinfo {author} {\bibfnamefont {Y.}~\bibnamefont
  {Zeng}}, \bibinfo {author} {\bibfnamefont {A.}~\bibnamefont {Hamma}}, \ and\
  \bibinfo {author} {\bibfnamefont {H.}~\bibnamefont {Fan}},\ }\href {\doibase
  10.1103/PhysRevB.94.125104} {\bibfield  {journal} {\bibinfo  {journal} {Phys.
  Rev. B}\ }\textbf {\bibinfo {volume} {94}},\ \bibinfo {pages} {125104}
  (\bibinfo {year} {2016})}\BibitemShut {NoStop}%
\bibitem [{\citenamefont {Chamon}(2005)}]{Chamon:2005}%
  \BibitemOpen
  \bibfield  {author} {\bibinfo {author} {\bibfnamefont {C.}~\bibnamefont
  {Chamon}},\ }\href {\doibase 10.1103/PhysRevLett.94.040402} {\bibfield
  {journal} {\bibinfo  {journal} {Phys. Rev. Lett.}\ }\textbf {\bibinfo
  {volume} {94}},\ \bibinfo {pages} {040402} (\bibinfo {year}
  {2005})}\BibitemShut {NoStop}%
\bibitem [{\citenamefont {Kim}\ and\ \citenamefont {Haah}(2016)}]{Kim:2016}%
  \BibitemOpen
  \bibfield  {author} {\bibinfo {author} {\bibfnamefont {I.~H.}\ \bibnamefont
  {Kim}}\ and\ \bibinfo {author} {\bibfnamefont {J.}~\bibnamefont {Haah}},\
  }\href {\doibase 10.1103/PhysRevLett.116.027202} {\bibfield  {journal}
  {\bibinfo  {journal} {Phys. Rev. Lett.}\ }\textbf {\bibinfo {volume} {116}},\
  \bibinfo {pages} {027202} (\bibinfo {year} {2016})}\BibitemShut {NoStop}%
\bibitem [{\citenamefont {Alicki}\ \emph {et~al.}(2010)\citenamefont {Alicki},
  \citenamefont {Horodecki}, \citenamefont {Horodecki},\ and\ \citenamefont
  {Horodecki}}]{Alicki:2010}%
  \BibitemOpen
  \bibfield  {author} {\bibinfo {author} {\bibfnamefont {R.}~\bibnamefont
  {Alicki}}, \bibinfo {author} {\bibfnamefont {M.}~\bibnamefont {Horodecki}},
  \bibinfo {author} {\bibfnamefont {P.}~\bibnamefont {Horodecki}}, \ and\
  \bibinfo {author} {\bibfnamefont {R.}~\bibnamefont {Horodecki}},\ }\href
  {\doibase 10.1142/S1230161210000023} {\bibfield  {journal} {\bibinfo
  {journal} {Open Systems \& Information Dynamics}\ }\textbf {\bibinfo {volume}
  {17}},\ \bibinfo {pages} {1} (\bibinfo {year} {2010})}\BibitemShut {NoStop}%
\bibitem [{\citenamefont {Maz\'{a}\v{c}}\ and\ \citenamefont
  {Hamma}(2012)}]{Mazac:2012}%
  \BibitemOpen
  \bibfield  {author} {\bibinfo {author} {\bibfnamefont {D.}~\bibnamefont
  {Maz\'{a}\v{c}}}\ and\ \bibinfo {author} {\bibfnamefont {A.}~\bibnamefont
  {Hamma}},\ }\href
  {http://www.sciencedirect.com/science/article/pii/S0003491612000723?via%3Dihub}
  {\bibfield  {journal} {\bibinfo  {journal} {Annals of Physics}\ }\textbf
  {\bibinfo {volume} {327}},\ \bibinfo {pages} {2096} (\bibinfo {year}
  {2012})}\BibitemShut {NoStop}%
\bibitem [{\citenamefont {Brown}\ \emph {et~al.}(2014)\citenamefont {Brown},
  \citenamefont {Al-Shimary},\ and\ \citenamefont {Pachos}}]{Brown:2014}%
  \BibitemOpen
  \bibfield  {author} {\bibinfo {author} {\bibfnamefont {B.~J.}\ \bibnamefont
  {Brown}}, \bibinfo {author} {\bibfnamefont {A.}~\bibnamefont {Al-Shimary}}, \
  and\ \bibinfo {author} {\bibfnamefont {J.~K.}\ \bibnamefont {Pachos}},\
  }\href {\doibase 10.1103/PhysRevLett.112.120503} {\bibfield  {journal}
  {\bibinfo  {journal} {Phys. Rev. Lett.}\ }\textbf {\bibinfo {volume} {112}},\
  \bibinfo {pages} {120503} (\bibinfo {year} {2014})}\BibitemShut {NoStop}%
\bibitem [{\citenamefont {Hamma}\ \emph {et~al.}(2009)\citenamefont {Hamma},
  \citenamefont {Castelnovo},\ and\ \citenamefont {Chamon}}]{Hamma:2009}%
  \BibitemOpen
  \bibfield  {author} {\bibinfo {author} {\bibfnamefont {A.}~\bibnamefont
  {Hamma}}, \bibinfo {author} {\bibfnamefont {C.}~\bibnamefont {Castelnovo}}, \
  and\ \bibinfo {author} {\bibfnamefont {C.}~\bibnamefont {Chamon}},\ }\href
  {\doibase 10.1103/PhysRevB.79.245122} {\bibfield  {journal} {\bibinfo
  {journal} {Phys. Rev. B}\ }\textbf {\bibinfo {volume} {79}},\ \bibinfo
  {pages} {245122} (\bibinfo {year} {2009})}\BibitemShut {NoStop}%
\bibitem [{\citenamefont {Pedrocchi}\ \emph {et~al.}(2013)\citenamefont
  {Pedrocchi}, \citenamefont {Hutter}, \citenamefont {Wootton},\ and\
  \citenamefont {Loss}}]{Pedrocchi:2013}%
  \BibitemOpen
  \bibfield  {author} {\bibinfo {author} {\bibfnamefont {F.~L.}\ \bibnamefont
  {Pedrocchi}}, \bibinfo {author} {\bibfnamefont {A.}~\bibnamefont {Hutter}},
  \bibinfo {author} {\bibfnamefont {J.~R.}\ \bibnamefont {Wootton}}, \ and\
  \bibinfo {author} {\bibfnamefont {D.}~\bibnamefont {Loss}},\ }\href {\doibase
  10.1103/PhysRevA.88.062313} {\bibfield  {journal} {\bibinfo  {journal} {Phys.
  Rev. A}\ }\textbf {\bibinfo {volume} {88}},\ \bibinfo {pages} {062313}
  (\bibinfo {year} {2013})}\BibitemShut {NoStop}%
\bibitem [{\citenamefont {Landon-Cardinal}\ \emph {et~al.}(2015)\citenamefont
  {Landon-Cardinal}, \citenamefont {Yoshida}, \citenamefont {Poulin},\ and\
  \citenamefont {Preskill}}]{Cardinal:2015}%
  \BibitemOpen
  \bibfield  {author} {\bibinfo {author} {\bibfnamefont {O.}~\bibnamefont
  {Landon-Cardinal}}, \bibinfo {author} {\bibfnamefont {B.}~\bibnamefont
  {Yoshida}}, \bibinfo {author} {\bibfnamefont {D.}~\bibnamefont {Poulin}}, \
  and\ \bibinfo {author} {\bibfnamefont {J.}~\bibnamefont {Preskill}},\ }\href
  {\doibase 10.1103/PhysRevA.91.032303} {\bibfield  {journal} {\bibinfo
  {journal} {Phys. Rev. A}\ }\textbf {\bibinfo {volume} {91}},\ \bibinfo
  {pages} {032303} (\bibinfo {year} {2015})}\BibitemShut {NoStop}%
\bibitem [{\citenamefont {Tsomokos}\ \emph {et~al.}(2011)\citenamefont
  {Tsomokos}, \citenamefont {Osborne},\ and\ \citenamefont
  {Castelnovo}}]{Tsomokos:2011}%
  \BibitemOpen
  \bibfield  {author} {\bibinfo {author} {\bibfnamefont {D.~I.}\ \bibnamefont
  {Tsomokos}}, \bibinfo {author} {\bibfnamefont {T.~J.}\ \bibnamefont
  {Osborne}}, \ and\ \bibinfo {author} {\bibfnamefont {C.}~\bibnamefont
  {Castelnovo}},\ }\href {\doibase 10.1103/PhysRevB.83.075124} {\bibfield
  {journal} {\bibinfo  {journal} {Phys. Rev. B}\ }\textbf {\bibinfo {volume}
  {83}},\ \bibinfo {pages} {075124} (\bibinfo {year} {2011})}\BibitemShut
  {NoStop}%
\bibitem [{\citenamefont {Wootton}\ and\ \citenamefont
  {Pachos}(2011)}]{Wootton:2011}%
  \BibitemOpen
  \bibfield  {author} {\bibinfo {author} {\bibfnamefont {J.~R.}\ \bibnamefont
  {Wootton}}\ and\ \bibinfo {author} {\bibfnamefont {J.~K.}\ \bibnamefont
  {Pachos}},\ }\href {\doibase 10.1103/PhysRevLett.107.030503} {\bibfield
  {journal} {\bibinfo  {journal} {Phys. Rev. Lett.}\ }\textbf {\bibinfo
  {volume} {107}},\ \bibinfo {pages} {030503} (\bibinfo {year}
  {2011})}\BibitemShut {NoStop}%
\bibitem [{\citenamefont {Stark}\ \emph {et~al.}(2011)\citenamefont {Stark},
  \citenamefont {Pollet}, \citenamefont {Imamo\ifmmode~\breve{g}\else
  \u{g}\fi{}lu},\ and\ \citenamefont {Renner}}]{Stark:2011}%
  \BibitemOpen
  \bibfield  {author} {\bibinfo {author} {\bibfnamefont {C.}~\bibnamefont
  {Stark}}, \bibinfo {author} {\bibfnamefont {L.}~\bibnamefont {Pollet}},
  \bibinfo {author} {\bibfnamefont {A.~m.~c.}\ \bibnamefont
  {Imamo\ifmmode~\breve{g}\else \u{g}\fi{}lu}}, \ and\ \bibinfo {author}
  {\bibfnamefont {R.}~\bibnamefont {Renner}},\ }\href {\doibase
  10.1103/PhysRevLett.107.030504} {\bibfield  {journal} {\bibinfo  {journal}
  {Phys. Rev. Lett.}\ }\textbf {\bibinfo {volume} {107}},\ \bibinfo {pages}
  {030504} (\bibinfo {year} {2011})}\BibitemShut {NoStop}%
\bibitem [{sup()}]{supp_info}%
  \BibitemOpen
  \href@noop {} {}\bibinfo {note} {See Supplementary material.}\BibitemShut
  {Stop}%
\bibitem [{\citenamefont {Stinchcombe}(1981)}]{Stinchcombe:1981}%
  \BibitemOpen
  \bibfield  {author} {\bibinfo {author} {\bibfnamefont {R.~B.}\ \bibnamefont
  {Stinchcombe}},\ }\href {http://stacks.iop.org/0022-3719/14/i=10/a=003}
  {\bibfield  {journal} {\bibinfo  {journal} {Journal of Physics C: Solid State
  Physics}\ }\textbf {\bibinfo {volume} {14}},\ \bibinfo {pages} {L263}
  (\bibinfo {year} {1981})}\BibitemShut {NoStop}%
\bibitem [{com()}]{comment}%
  \BibitemOpen
  \href@noop {} {}\bibinfo {note} {Generally, in a stabilizer Hamiltonian, one
  deals with two concepts of mass, i.e. effective mass and mass gap. The latter
  is defined as energy needed for creation/anihilation of anyons and controlled
  by $j_e$ and $j_m$ parameters.}\BibitemShut {Stop}%
\bibitem [{\citenamefont {Balay}\ \emph {et~al.}(2016)\citenamefont {Balay},
  \citenamefont {Abhyankar}, \citenamefont {Adams}, \citenamefont {Brune},
  \citenamefont {Buschelman}, \citenamefont {Dalcin}, \citenamefont {Gropp},
  \citenamefont {Smith}, \citenamefont {Karpeyev}, \citenamefont {Kaushik}
  \emph {et~al.}}]{petsc-user-ref}%
  \BibitemOpen
  \bibfield  {author} {\bibinfo {author} {\bibfnamefont {S.}~\bibnamefont
  {Balay}}, \bibinfo {author} {\bibfnamefont {S.}~\bibnamefont {Abhyankar}},
  \bibinfo {author} {\bibfnamefont {M.}~\bibnamefont {Adams}}, \bibinfo
  {author} {\bibfnamefont {P.}~\bibnamefont {Brune}}, \bibinfo {author}
  {\bibfnamefont {K.}~\bibnamefont {Buschelman}}, \bibinfo {author}
  {\bibfnamefont {L.}~\bibnamefont {Dalcin}}, \bibinfo {author} {\bibfnamefont
  {W.}~\bibnamefont {Gropp}}, \bibinfo {author} {\bibfnamefont
  {B.}~\bibnamefont {Smith}}, \bibinfo {author} {\bibfnamefont
  {D.}~\bibnamefont {Karpeyev}}, \bibinfo {author} {\bibfnamefont
  {D.}~\bibnamefont {Kaushik}},  \emph {et~al.},\ }\href@noop {} {\emph
  {\bibinfo {title} {Petsc users manual revision 3.7}}},\ \bibinfo {type}
  {Tech. Rep.}\ (\bibinfo  {institution} {Argonne National Lab.(ANL), Argonne,
  IL (United States)},\ \bibinfo {year} {2016})\BibitemShut {NoStop}%
\bibitem [{\citenamefont {Balay}\ \emph {et~al.}(1997)\citenamefont {Balay},
  \citenamefont {Gropp}, \citenamefont {McInnes},\ and\ \citenamefont
  {Smith}}]{petsc-efficient}%
  \BibitemOpen
  \bibfield  {author} {\bibinfo {author} {\bibfnamefont {S.}~\bibnamefont
  {Balay}}, \bibinfo {author} {\bibfnamefont {W.~D.}\ \bibnamefont {Gropp}},
  \bibinfo {author} {\bibfnamefont {L.~C.}\ \bibnamefont {McInnes}}, \ and\
  \bibinfo {author} {\bibfnamefont {B.~F.}\ \bibnamefont {Smith}},\ }in\
  \href@noop {} {\emph {\bibinfo {booktitle} {Modern Software Tools in
  Scientific Computing}}},\ \bibinfo {editor} {edited by\ \bibinfo {editor}
  {\bibfnamefont {E.}~\bibnamefont {Arge}}, \bibinfo {editor} {\bibfnamefont
  {A.~M.}\ \bibnamefont {Bruaset}}, \ and\ \bibinfo {editor} {\bibfnamefont
  {H.~P.}\ \bibnamefont {Langtangen}}}\ (\bibinfo  {publisher}
  {Birkh{\"{a}}user Press},\ \bibinfo {year} {1997})\ pp.\ \bibinfo {pages}
  {163--202}\BibitemShut {NoStop}%
\bibitem [{\citenamefont {Hernandez}\ \emph {et~al.}(2005)\citenamefont
  {Hernandez}, \citenamefont {Roman},\ and\ \citenamefont
  {Vidal}}]{Hernandez:2005}%
  \BibitemOpen
  \bibfield  {author} {\bibinfo {author} {\bibfnamefont {V.}~\bibnamefont
  {Hernandez}}, \bibinfo {author} {\bibfnamefont {J.~E.}\ \bibnamefont
  {Roman}}, \ and\ \bibinfo {author} {\bibfnamefont {V.}~\bibnamefont
  {Vidal}},\ }\href {\doibase 10.1145/1089014.1089019} {\bibfield  {journal}
  {\bibinfo  {journal} {ACM Trans. Math. Softw.}\ }\textbf {\bibinfo {volume}
  {31}},\ \bibinfo {pages} {351} (\bibinfo {year} {2005})}\BibitemShut
  {NoStop}%
\bibitem [{\citenamefont {Page}(1993)}]{Page:1993}%
  \BibitemOpen
  \bibfield  {author} {\bibinfo {author} {\bibfnamefont {D.~N.}\ \bibnamefont
  {Page}},\ }\href {\doibase 10.1103/PhysRevLett.71.1291} {\bibfield  {journal}
  {\bibinfo  {journal} {Phys. Rev. Lett.}\ }\textbf {\bibinfo {volume} {71}},\
  \bibinfo {pages} {1291} (\bibinfo {year} {1993})}\BibitemShut {NoStop}%
\bibitem [{foo()}]{footnote}%
  \BibitemOpen
  \href@noop {} {}\bibinfo {note} {In this setting, we investigate the
  translationally invariant pre-quench states and the observed non-ergodic
  behavior takes place even at the moderate value of $t_m$.}\BibitemShut
  {Stop}%
\bibitem [{\citenamefont {Grover}\ and\ \citenamefont
  {Fisher}(2014)}]{Grover:2014}%
  \BibitemOpen
  \bibfield  {author} {\bibinfo {author} {\bibfnamefont {T.}~\bibnamefont
  {Grover}}\ and\ \bibinfo {author} {\bibfnamefont {M.~P.~A.}\ \bibnamefont
  {Fisher}},\ }\href {http://stacks.iop.org/1742-5468/2014/i=10/a=P10010}
  {\bibfield  {journal} {\bibinfo  {journal} {Journal of Statistical Mechanics:
  Theory and Experiment}\ }\textbf {\bibinfo {volume} {2014}},\ \bibinfo
  {pages} {P10010} (\bibinfo {year} {2014})}\BibitemShut {NoStop}%
\bibitem [{\citenamefont {Levin}\ and\ \citenamefont {Wen}(2005)}]{Levin:2005}%
  \BibitemOpen
  \bibfield  {author} {\bibinfo {author} {\bibfnamefont {M.~A.}\ \bibnamefont
  {Levin}}\ and\ \bibinfo {author} {\bibfnamefont {X.-G.}\ \bibnamefont
  {Wen}},\ }\href {\doibase 10.1103/PhysRevB.71.045110} {\bibfield  {journal}
  {\bibinfo  {journal} {Phys. Rev. B}\ }\textbf {\bibinfo {volume} {71}},\
  \bibinfo {pages} {045110} (\bibinfo {year} {2005})}\BibitemShut {NoStop}%
\bibitem [{\citenamefont {Garrison}\ \emph {et~al.}(2017)\citenamefont
  {Garrison}, \citenamefont {Mishmash},\ and\ \citenamefont
  {Fisher}}]{Garrison:2016}%
  \BibitemOpen
  \bibfield  {author} {\bibinfo {author} {\bibfnamefont {J.~R.}\ \bibnamefont
  {Garrison}}, \bibinfo {author} {\bibfnamefont {R.~V.}\ \bibnamefont
  {Mishmash}}, \ and\ \bibinfo {author} {\bibfnamefont {M.~P.~A.}\ \bibnamefont
  {Fisher}},\ }\href {\doibase 10.1103/PhysRevB.95.054204} {\bibfield
  {journal} {\bibinfo  {journal} {Phys. Rev. B}\ }\textbf {\bibinfo {volume}
  {95}},\ \bibinfo {pages} {054204} (\bibinfo {year} {2017})}\BibitemShut
  {NoStop}%
\bibitem [{\citenamefont {{Veness}}\ \emph {et~al.}(2016)\citenamefont
  {{Veness}}, \citenamefont {{Essler}},\ and\ \citenamefont
  {{Fisher}}}]{Veness:2016}%
  \BibitemOpen
  \bibfield  {author} {\bibinfo {author} {\bibfnamefont {T.}~\bibnamefont
  {{Veness}}}, \bibinfo {author} {\bibfnamefont {F.~H.~L.}\ \bibnamefont
  {{Essler}}}, \ and\ \bibinfo {author} {\bibfnamefont {M.~P.~A.}\ \bibnamefont
  {{Fisher}}},\ }\href@noop {} {\bibfield  {journal} {\bibinfo  {journal}
  {ArXiv e-prints}\ } (\bibinfo {year} {2016})},\ \Eprint
  {http://arxiv.org/abs/1611.02075} {arXiv:1611.02075 [cond-mat.stat-mech]}
  \BibitemShut {NoStop}%
\end{thebibliography}

\begin{thebibliography}{11}%
\makeatletter
\providecommand \@ifxundefined [1]{%
 \@ifx{#1\undefined}
}%
\providecommand \@ifnum [1]{%
 \ifnum #1\expandafter \@firstoftwo
 \else \expandafter \@secondoftwo
 \fi
}%
\providecommand \@ifx [1]{%
 \ifx #1\expandafter \@firstoftwo
 \else \expandafter \@secondoftwo
 \fi
}%
\providecommand \natexlab [1]{#1}%
\providecommand \enquote  [1]{``#1''}%
\providecommand \bibnamefont  [1]{#1}%
\providecommand \bibfnamefont [1]{#1}%
\providecommand \citenamefont [1]{#1}%
\providecommand \href@noop [0]{\@secondoftwo}%
\providecommand \href [0]{\begingroup \@sanitize@url \@href}%
\providecommand \@href[1]{\@@startlink{#1}\@@href}%
\providecommand \@@href[1]{\endgroup#1\@@endlink}%
\providecommand \@sanitize@url [0]{\catcode `\\12\catcode `\$12\catcode
  `\&12\catcode `\#12\catcode `\^12\catcode `\_12\catcode `\%12\relax}%
\providecommand \@@startlink[1]{}%
\providecommand \@@endlink[0]{}%
\providecommand \url  [0]{\begingroup\@sanitize@url \@url }%
\providecommand \@url [1]{\endgroup\@href {#1}{\urlprefix }}%
\providecommand \urlprefix  [0]{URL }%
\providecommand \Eprint [0]{\href }%
\providecommand \doibase [0]{http://dx.doi.org/}%
\providecommand \selectlanguage [0]{\@gobble}%
\providecommand \bibinfo  [0]{\@secondoftwo}%
\providecommand \bibfield  [0]{\@secondoftwo}%
\providecommand \translation [1]{[#1]}%
\providecommand \BibitemOpen [0]{}%
\providecommand \bibitemStop [0]{}%
\providecommand \bibitemNoStop [0]{.\EOS\space}%
\providecommand \EOS [0]{\spacefactor3000\relax}%
\providecommand \BibitemShut  [1]{\csname bibitem#1\endcsname}%
\let\auto@bib@innerbib\@empty
\bibitem [{\citenamefont {Amestoy}\ \emph {et~al.}(2006)\citenamefont
  {Amestoy}, \citenamefont {Guermouche}, \citenamefont {L'Excellent},\ and\
  \citenamefont {Pralet}}]{MUMPS:2006}%
  \BibitemOpen
  \bibfield  {author} {\bibinfo {author} {\bibfnamefont {P.~R.}\ \bibnamefont
  {Amestoy}}, \bibinfo {author} {\bibfnamefont {A.}~\bibnamefont {Guermouche}},
  \bibinfo {author} {\bibfnamefont {J.-Y.}\ \bibnamefont {L'Excellent}}, \ and\
  \bibinfo {author} {\bibfnamefont {S.}~\bibnamefont {Pralet}},\ }\href@noop {}
  {\bibfield  {journal} {\bibinfo  {journal} {Parallel Computing}\ }\textbf
  {\bibinfo {volume} {32}},\ \bibinfo {pages} {136} (\bibinfo {year}
  {2006})}\BibitemShut {NoStop}%
\bibitem [{\citenamefont {Or\'us}\ and\ \citenamefont
  {Vidal}(2008)}]{Orus:2008}%
  \BibitemOpen
  \bibfield  {author} {\bibinfo {author} {\bibfnamefont {R.}~\bibnamefont
  {Or\'us}}\ and\ \bibinfo {author} {\bibfnamefont {G.}~\bibnamefont {Vidal}},\
  }\href {\doibase 10.1103/PhysRevB.78.155117} {\bibfield  {journal} {\bibinfo
  {journal} {Phys. Rev. B}\ }\textbf {\bibinfo {volume} {78}},\ \bibinfo
  {pages} {155117} (\bibinfo {year} {2008})}\BibitemShut {NoStop}%
\bibitem [{\citenamefont {Pollmann}\ \emph {et~al.}(2010)\citenamefont
  {Pollmann}, \citenamefont {Turner}, \citenamefont {Berg},\ and\ \citenamefont
  {Oshikawa}}]{Pollmann:2010}%
  \BibitemOpen
  \bibfield  {author} {\bibinfo {author} {\bibfnamefont {F.}~\bibnamefont
  {Pollmann}}, \bibinfo {author} {\bibfnamefont {A.~M.}\ \bibnamefont
  {Turner}}, \bibinfo {author} {\bibfnamefont {E.}~\bibnamefont {Berg}}, \ and\
  \bibinfo {author} {\bibfnamefont {M.}~\bibnamefont {Oshikawa}},\ }\href
  {\doibase 10.1103/PhysRevB.81.064439} {\bibfield  {journal} {\bibinfo
  {journal} {Phys. Rev. B}\ }\textbf {\bibinfo {volume} {81}},\ \bibinfo
  {pages} {064439} (\bibinfo {year} {2010})}\BibitemShut {NoStop}%
\bibitem [{\citenamefont {Pollmann}\ and\ \citenamefont
  {Turner}(2012)}]{Pollmann:2012}%
  \BibitemOpen
  \bibfield  {author} {\bibinfo {author} {\bibfnamefont {F.}~\bibnamefont
  {Pollmann}}\ and\ \bibinfo {author} {\bibfnamefont {A.~M.}\ \bibnamefont
  {Turner}},\ }\href {\doibase 10.1103/PhysRevB.86.125441} {\bibfield
  {journal} {\bibinfo  {journal} {Phys. Rev. B}\ }\textbf {\bibinfo {volume}
  {86}},\ \bibinfo {pages} {125441} (\bibinfo {year} {2012})}\BibitemShut
  {NoStop}%
\end{thebibliography}
\end{document}